\documentclass[10pt]{iopart}

\eqnobysec

\usepackage{iopams}
\usepackage{amstext}
\usepackage{placeins}

\begin{document}

\title[Metric variations in higher-derivative
gravity and stress-energy tensor] {Irreducible forms for the metric
variations of the action terms of sixth-order gravity and
approximated stress-energy tensor}

\author{Yves D\'ecanini and Antoine Folacci}
\address{ UMR CNRS 6134 SPE, \\ Equipe
Physique Semi-Classique (et) de la Mati\`ere Condens\'ee,
\\ Universit\'e de Corse, Facult\'e des Sciences, Bo{\^\i}te
Postale 52, 20250 Corte, France}
\eads{\mailto{decanini@univ-corse.fr},
\mailto{folacci@univ-corse.fr}}

\begin{abstract}
\noindent We provide irreducible expressions for the metric
variations of the gravitational action terms constructed from the 17
curvature invariants of order six in derivatives of the metric
tensor i.e. from the geometrical terms appearing in the diagonal
heat-kernel or Gilkey-DeWitt coefficient $a_3$. We then express, for
a four dimensional spacetime, the approximated stress-energy tensor
constructed from the renormalized DeWitt-Schwinger effective action
associated with a massive scalar field. We also construct, for
higher dimensional spacetimes, the infinite counterterms of order
six in derivatives of the metric tensor appearing in the left hand
side of Einstein equations as well as the contribution associated
with the cubic Lovelock gravitational action. In an appendix, we
provide a list of geometrical relations we have used and which are
more generally helpful for calculations in two-loop quantum gravity
in a four dimensional background or for calculations in one-loop
quantum gravity in higher dimensional background. We also obtain the
approximated stress-energy tensors associated with a massive spinor
field and a massive vector field propagating in a four dimensional
background.
\end{abstract}



\section{Introduction}

Calculations which must be carried out in field theories defined on
curved spacetimes are in general highly non-trivial. This is more
particularly true in the context of renormalization of quantum
fields and of quantum gravity but, even at the classical level,
analogous difficulties can appear, for example, in the context of
the radiation reaction problem of gravitational wave theory. This is
mainly due to the systematic occurrence, in these calculations, of
Riemann polynomials (i.e., polynomials formed from the Riemann
tensor by covariant differentiation, multiplication and contraction)
whose complexity, degree and number rapidly increase with the
precision of the approximations needed or with the dimension of the
gravitational background considered. The results of these
calculations may be moreover darkened because of the non-uniqueness
of their final forms what moreover complicates the comparison
between the works completed by different authors: Indeed, the
symmetries of the Riemann tensor as well as Bianchi identities are
not used in a uniform manner and monomials formed from the Riemann
tensor may be linearly dependent in non-trivial ways. In a beautiful
and very useful article \cite{FKWC1992}, Fulling, King, Wybourne and
Cummings (FKWC) have proposed to cure this last problem by expanding
systematically the Riemann polynomials encountered in calculations
on standard bases constructed from group theoretical considerations.
They have also displayed such bases for scalar Riemann polynomials
of order eight or less in derivatives of the metric tensor and for
tensorial Riemann polynomials of order six or less.

\bigskip

In Section 2 of the present article, we shall use FKWC-results to
provide irreducible expressions for the metric variations (i.e. for
the functional derivatives with respect to the metric tensor) of the
action terms associated with the 17 basis elements for the scalar
Riemann polynomials of order six in derivatives of the metric tensor
(the so-called curvature invariants of order six). We also provide
the irreducible expression for the cubic Lovelock tensor, i.e. the
metric variation of the cubic Lovelock gravitational action
\cite{Lovelock71,LovelockRund}. The elements of the scalar
FKWC-basis previously considered appear in the expression of the
diagonal heat-kernel (or Gilkey \cite{Sakai71,Gilkey75,Gilkey84})
coefficient $a_3$ and therefore in the unrenormalized or
renormalized DeWitt-Schwinger effective action
\cite{Schwinger1951,DeWitt65,BirrellDavies,Avramidi_PhD,AvramidiLNP2000,DeWitt03}
associated with massive fields propagating in curved spacetime. As a
consequence, our results will permit us to unambiguously express, in
Section 3, for a four dimensional spacetime, the approximated
stress-energy tensor constructed from the renormalized effective
action associated with a massive scalar field as well as to
construct, for an arbitrary six dimensional spacetime, the infinite
counterterms of order six in derivatives of the metric tensor which
appear in the left hand side of the bare Einstein equations. In a
brief conclusion (Section 4), we shall consider some possible
immediate prolongations of our present work. Finally, in an
appendix, we shall provide a list of geometrical relations we have
used (these relations are more generally helpful for calculations in
two-loop quantum gravity in a four dimensional background or for
calculations in one-loop quantum gravity in higher dimensional
background) and we shall briefly extend the result obtained in
Section 3 by providing, in the large mass limit, simplified
expressions for the approximated stress-energy tensors associated
with a massive spinor field and a massive vector field propagating
in a four dimensional background.

\bigskip

All our results are obtained by using the geometrical conventions of
Hawking and Ellis \cite{HawkingEllis} as far as the definitions of
the scalar curvature $R$, the Ricci tensor $R_{pq}$ and the Riemann
tensor $R_{pqrs}$ are concerned and the commutation of covariant
derivatives in the form
\begin{equation}\label{CD_NabNabTensor}
T^{p \dots}_{\phantom{p} q \dots ;sr} - T^{p \dots}_{\phantom{p} q
\dots ;rs} =  + R^{p}_{\phantom{p} t rs} T^{t \dots}_{\phantom{t} q
\dots } + \dots - R^{t}_{\phantom{t} q rs} T^{p \dots}_{\phantom{p}
t \dots} - \dots
\end{equation}
We use furthermore the FKWC-notations ${\cal R}^r_{s,q}$ and ${\cal
R}^r_{\lbrace{\lambda_1 \dots \rbrace}}$: ${\cal R}^r_{s,q}$ denotes
the space of Riemann polynomials of rank r (number of free indices),
order s (number of differentiation of the metric tensor) and degree
q (number of factors $\nabla^q R_{\dots}^{\dots}$) while ${\cal
R}^r_{\lbrace{\lambda_1 \dots \rbrace}}$ denotes the space of
Riemann polynomials of rank r spanned by contractions of products of
the type $\nabla^{\lambda_1}R_{\dots}^{\dots}$. We refer to the
FKWC-article \cite{FKWC1992} for more precisions on these notations
and rigor on the subject.

\section{Irreducible forms for the metric variations of the action terms
constructed from 17 curvature invariants of order six in derivatives
of the metric tensor and for the cubic Lovelock tensor}

In this section, we consider spacetime as an arbitrary
$D$-dimensional pseudo-Riemannian manifold $({\cal M},g_{ab})$ and
we assume that its dimension $D$ is sufficiently high in order to
avoid any degeneracy of the Riemann tensor as well as any
topological constraint associated with Euler-Gauss-Bonnet-Lovelock
densities and numbers. Later, we shall drop this strong hypothesis.

\subsection{FKWC-basis for Riemann polynomials of order six and rank zero}

The most general expression of a gravitational lagrangian of order
six in derivatives of the metric tensor is obtained by expanding it
on the FKWC-basis for Riemann polynomials of order six and rank
zero. This basis consists of the 17 following elements
\cite{FKWC1992}:
\begin{eqnarray} \label{scalar_O6}
&  {\cal R}^0_{6,1}: & \Box \Box R     \nonumber \\
&  {\cal R}^0_{\lbrace{2,0\rbrace}}: & R\Box R \quad   R_{;p q}
R^{pq} \quad R_{pq} \Box R^{pq} \quad R_{pq ; rs}R^{prqs} \nonumber \\
&   {\cal R}^0_{\lbrace{1,1 \rbrace}}: & R_{;p}R^{;p} \quad R_{pq;r}
R^{pq;r} \quad R_{pq;r} R^{pr;q} \quad R_{pqrs;t} R^{pqrs;t}
    \nonumber \\
&   {\cal R}^0_{6,3}: &  R^3   \quad  RR_{pq} R^{pq}  \quad R_{pq}
R^{p}_{\phantom{p} r}R^{qr}  \quad R_{pq}R_{rs}R^{prqs} \quad
RR_{pqrs} R^{pqrs}
 \nonumber \\
&   &  R_{pq}R^p_{\phantom{p} rst} R^{qrst } \quad R_{pqrs}R^{pquv}
R^{rs}_{\phantom{rs} uv} \quad R_{prqs} R^{p \phantom{u}
q}_{\phantom{p} u \phantom{q} v} R^{r u s v}.
\end{eqnarray}

\bigskip

This basis is a natural one and is often used on this form in
literature. However, it should be noted that certain authors prefer
to use the scalar monomial $R_{pqrs} \Box R^{pqrs}$ instead of the
scalar monomial $R_{pq ; rs}R^{prqs}$. This is the case of Gilkey in
Refs.~\cite{Gilkey75,Gilkey84}. This choice is only a matter of
taste because these two terms appear equally in the calculations
carried out in field theories defined on curved spacetimes and the
elimination of one of them can be achieved by using the identity
(\ref{scalar_O6_comp1}). It should be also noted that in these
calculations, other Riemann monomials of order six and rank zero
such as $R_{pq}R^{pr}_{\phantom{pr};qr}$, $R^{prqs}R_{pquv}R_{rs
\phantom{uv}}^{\phantom{rs} uv}$, $R_{pq}^{\phantom{pq}
rs}R^{puqv}R_{rusv}$ and $R_{prqs}R^{puqv}R_{\phantom{r} v
\phantom{s} u}^{r \phantom{u} s \phantom{v} }$ are systematically
encountered. They can be eliminated by using the geometrical
identities (\ref{scalar_O6_comp0a}) and
(\ref{scalar_O6_RCub1})-(\ref{scalar_O6_RCub3}) which permit us to
expand them on the FKWC-basis (\ref{scalar_O6}).

\subsection{FKWC-basis for Riemann polynomials of order six and rank two}

The functional derivatives with respect to the metric tensor of an
action term constructed from a gravitational lagrangian of order six
is obtained by expanding it on the FKWC-basis for Riemann
polynomials of order six and rank two. This basis consists of the 42
following elements \cite{FKWC1992}:
\begin{eqnarray}\label{Tensor2_O6}
&    {\cal R}^2_{6,1}: &  (\Box R)_{;ab} \quad  \Box \Box R_{a b} \nonumber \\
&  {\cal R}^2_{\lbrace{2,0\rbrace}}:  & R R_{;a b} \quad  (\Box R)
R_{a b} \quad   R_{;p a} R^{p}_{\phantom{p}b} \quad R \Box R_{a b}
\quad R_{p a} \Box R^p_{\phantom{p} b} \nonumber \\
& &  R^{pq} R_{pq;a b} \quad
 R^{pq} R_{p a ; bq}  \quad R^{pq} R_{ab; pq} \quad R^{;pq}R_{p
a q b} \nonumber \\
& &   (\Box R^{pq})R_{p a q b} \quad R^{pq;r}_{\phantom{pq;r} a}
R_{rqp b} \quad R^{p \phantom{a}; qr}_{\phantom{p } a} R_{pqr  b}
\quad R^{pqrs} R_{pqrs ; a b
} \nonumber \\
&  {\cal R}^2_{\lbrace{1,1 \rbrace}}:  & R_{;a} R_{;b} \quad R_{;p }
R^p_{\phantom{p} a;b} \quad R_{;p } R_{ab}^{\phantom{ab}; p} \quad
R^{pq}_{\phantom{pq};a}R_{pq;b}
   \quad R^{pq}_{\phantom{pq} ; a} R_{b p;q}  \nonumber \\
& & R^p_{\phantom{p} a;q} R_{p b}^{\phantom{p b};q} \quad
R^p_{\phantom{p} a;q} R^q_{\phantom{q} b;p}  \quad R^{pq;r} R_{rqp
a;b } \quad R^{pq;r} R_{p a q b;r} \nonumber \\
& & R^{pqrs}_{\phantom{pqrs};a} R_{pqrs ; b } \quad
R^{pqr}_{\phantom{pqr}a;s}R_{pqr b}^{\phantom{pqr
b};s}   \nonumber \\
&  {\cal R}^2_{6,3}: & R^2 R_{ab} \quad R R_{p a} R^p_{\phantom{p}b}
\quad R^{pq}R_{pq}R_{ab} \quad R^{pq}R_{p a}R_{q b} \quad R R^{pq}R_{p a q b} \nonumber \\
& &   R^{pr}R^q_{\phantom{q} r}R_{p a q b} \quad
R^{pq}R^r_{\phantom{r}  a}R_{ rqp  b }
 \quad R R^{pqr}_{\phantom{pqr}a }R_{pqr b} \nonumber \\
& & R_{ab}R^{pqrs} R_{pqrs } \quad R^p_{\phantom{p}
a}R^{qrs}_{\phantom{qrs} p }R_{ qrs  b } \quad
R^{pq}R^{rs}_{\phantom{rs} pa}R_{rs q b} \nonumber \\
& & R_{pq}R^{p r q s}R_{r a s b} \quad R_{pq}R^{p rs}_{\phantom{p
rs}a}R^q_{\phantom{q} rs b}\quad R^{pq rs}R_{pq t a }R_{rs
\phantom{t} b}^{\phantom{rs} t}
 \nonumber \\
& & R^{p r q s}R^t_{\phantom{t} pq a}R_{t rs b}  \quad
R^{pqr}_{\phantom{pqr} s } R_{pqr t}R^{s \phantom{a} t}_{\phantom{s}
a \phantom{t} b}.
\end{eqnarray}

\bigskip

Here, it should be noted that we have slightly modified the
FKWC-basis of Ref.~\cite{FKWC1992}:

\qquad i) We have replaced the term $R^{pq} R_{p a ; q b}$ proposed
in Ref.~\cite{FKWC1992} by the term $R^{pq} R_{p a ; b q}$. We think
it is more interesting to work with the latter which directly
reduces to an element of the scalar basis (\ref{scalar_O6}) by
contraction on the free indices $a$ and $b$. In fact, these two
terms are linked by the geometrical identity (\ref{tensor_O6_comp4})
so it is easy to return to the original FKWC-basis.

\qquad ii) We have replaced the terms $R^{pq;r}_{\phantom{pq;r} a}
R_{prq b}$, $R^{pq;r} R_{prq a;b}$ and $R^{pq}R^r_{\phantom{r}
a}R_{prq b}$ proposed in Ref.~\cite{FKWC1992} by their opposites
respectively given by $R^{pq;r}_{\phantom{pq;r} a} R_{rqp b}$,
$R^{pq;r} R_{rqp a;b}$ and $R^{pq}R^r_{\phantom{r} a}R_{rqp b}$.
This choice has been done for obvious mnemotechnic reasons.

\bigskip

In the calculations carried out in field theories defined on curved
spacetimes as well as in quantum gravity and more particularly in
the calculations we shall achieve in the present section, other
Riemann monomials of order six and rank two which are not in the
FKWC-basis (\ref{Tensor2_O6}) are systematically encountered. They
can be eliminated (i.e., expanded on the FKWC-basis
(\ref{Tensor2_O6})) more or less trivially from the geometrical
identities
(\ref{tensor_O6_comp2}),(\ref{tensor_O6_comp4}),(\ref{tensor_O6_comp5}),(\ref{tensor_O6_comp14}),
(\ref{tensor_O6_comp15a})-(\ref{tensor_O6_comp15a}),(\ref{tensor_O6_comp6}),(\ref{tensor_O6_comp7}),(\ref{tensor_O6_comp8}),(\ref{tensor_O6_comp11})
and (\ref{tensor_O6_comp12b})-(\ref{tensor_O6_comp12e}) we have
obtained and displayed in the Appendix.

\subsection{Action
terms constructed from the 17 scalar Riemann monomials of order six}

From the 17 elements of the FKWC-basis (\ref{scalar_O6}), we can
construct 17 action terms which permit us to express the most
general gravitational action constructed from a lagrangian of order
six in derivatives of the metric tensor. For a general spacetime
$({\cal M},g_{ab})$ of arbitrary dimension $D$ (with $D$
sufficiently high), these 17 action terms are independent but if we
assume that the considered spacetime has no boundary (i.e., if
$\partial {\cal M}= \emptyset $), some of these terms are linked
together due to Stokes's theorem. From now on, we shall work under
this hypothesis and therefore consider that, for any vector field
$V$, we have
\begin{equation}\label{StokesTh}
 \int_{\cal M} d^D x \sqrt{-g}~V^p_{\phantom{p} ;p} =0.
\end{equation}

 By using integration by part, the contracted Bianchi identities (\ref{AppBianchi_2a}) and
(\ref{AppBianchi_2b}) as well as the geometrical identities
(\ref{scalar_O6_comp0a}) and (\ref{scalar_O6_comp1}) and Stokes's
theorem in the form (\ref{StokesTh}), it is easy to prove that only
the ten action terms
\begin{eqnarray}
& \int_{\cal M} d^D x\sqrt{-g} ~R\Box R  \label{actionO6_1}\\
& \int_{\cal M} d^D x\sqrt{-g} ~R_{pq} \Box R^{pq} \\
& \int_{\cal M} d^D x\sqrt{-g} ~R^3 \\
& \int_{\cal M} d^D x\sqrt{-g}  ~RR_{pq} R^{pq} \\
& \int_{\cal M} d^D x\sqrt{-g}  ~R_{pq} R^{p}_{\phantom{p}
r}R^{qr} \\
& \int_{\cal M} d^D x\sqrt{-g}  ~R_{pq}R_{rs}R^{prqs} \\
& \int_{\cal M} d^D x\sqrt{-g}  ~RR_{pqrs} R^{pqrs} \\
& \int_{\cal M} d^D x\sqrt{-g}  ~R_{pq}R^p_{\phantom{p} rst} R^{qrst } \\
& \int_{\cal M} d^D x\sqrt{-g}  ~R_{pqrs}R^{pquv} R^{rs}_{\phantom{rs} uv}\\
& \int_{\cal M} d^D x\sqrt{-g}  ~R_{prqs} R^{p \phantom{u}
q}_{\phantom{p} u \phantom{q} v} R^{r u s v} \label{actionO6_10}
\end{eqnarray}
are independent and that one of the remaining action terms vanishes
while the six other ones can be expressed in terms of some of the
previous ones:
\begin{eqnarray}
& \int_{\cal M} d^D x \sqrt{-g} ~\Box \Box R= 0  \label{actionO6_11}\\
& \int_{\cal M} d^D x \sqrt{-g} ~R_{;p q}
R^{pq} =   \int_{\cal M} d^D x\sqrt{-g}  \left( \frac{1}{2} \, R\Box R \right) \label{actionO6_12}\\
& \int_{\cal M} d^D x \sqrt{-g} ~R_{pq ; rs}R^{prqs} = \nonumber
\\
& \quad \int_{\cal M} d^D x  \sqrt{-g} \left( -\frac{1}{4} \,  R\Box
R + R_{pq} \Box R^{pq}  - R_{pq} R^{p}_{\phantom{p}
r}R^{qr} + R_{pq}R_{rs}R^{prqs} \right) \nonumber \\
& \label{actionO6_13} \\
& \int_{\cal M} d^D x\sqrt{-g} ~R_{;p}R^{;p} =
\int_{\cal M} d^D x \sqrt{-g} \left(-R\Box R\right) \\
& \int_{\cal M} d^D x\sqrt{-g} ~R_{pq;r}
R^{pq;r} = \int_{\cal M} d^D x\sqrt{-g}  \left(-R_{pq} \Box R^{pq}\right)   \\
& \int_{\cal M} d^D x\sqrt{-g} ~R_{pq;r} R^{pr;q} = \nonumber
\\
& \quad \int_{\cal M} d^D x \sqrt{-g} \left( -\frac{1}{4} R\Box R
-R_{pq} R^{p}_{\phantom{p}
r}R^{qr} + R_{pq}R_{rs}R^{prqs} \right)\\
& \int_{\cal M} d^D x \sqrt{-g} ~R_{pqrs;t} R^{pqrs;t}= \nonumber
\\
& \quad \int_{\cal M} d^D x \sqrt{-g} \left(  R\Box R -4 R_{pq} \Box
R^{pq} + 4  R_{pq} R^{p}_{\phantom{p} r}R^{qr} - 4
R_{pq}R_{rs}R^{prqs} \right.\nonumber \\
& \qquad\quad \left. -2 R_{pq}R^p_{\phantom{p} rst} R^{qrst } +
R_{pqrs}R^{pquv} R^{rs}_{\phantom{rs} uv} + 4 R_{prqs} R^{p
\phantom{u} q}_{\phantom{p} u \phantom{q} v} R^{r u s v} \right).
\nonumber \\
&  \label{actionO6_17}
\end{eqnarray}

\subsection{Explicit irreducible expressions for the metric variations of the action terms
 constructed from the 17 scalar Riemann monomials of order six}

The functional derivatives with respect to the metric tensor of the
ten independent action terms (\ref{actionO6_1})-(\ref{actionO6_10})
can be obtained by using the behaviour of these action terms in the
variation
\begin{equation}\label{varTensMet1}
g_{\mu \nu } \to g_{\mu \nu } + \delta g_{\mu \nu }
\end{equation}
of the metric tensor. From the corresponding variations of the
geometrical tensors $\sqrt{-g}$, $R$, $\Box R$, $R_{ab}$, $\Box
R_{ab}$ and $R_{abcd}$ given in Subsection (5.2) of the Appendix
(see (\ref{varTensMet4a})-(\ref{varTensMet4g})) or in
Ref.~\cite{ChristensenBarth83} and after tedious calculations, we
have expanded the functional derivatives of the ten action terms
(\ref{actionO6_1})-(\ref{actionO6_10}) on the FKWC-basis
(\ref{scalar_O6}) and (\ref{Tensor2_O6}) and we have obtained the
following results:

\begin{eqnarray}\label{FD_06_1}
 H_{ab}^{\lbrace{2,0\rbrace}(1)} & \equiv \frac{1}{
\sqrt{-g}}\frac{\delta}{\delta g^{ab}} \int_{\cal M} d^D x\sqrt{-g}
~R\Box R  \nonumber \\
  & =  2\, (\Box R)_{;ab} -2 \,  (\Box R) R_{a b} + R_{;a}
R_{;b} \nonumber \\
& \qquad \quad + g_{ab} \left[ -2\, \Box \Box R    - (1/2) \,
R_{;p}R^{;p}\right],
\end{eqnarray}

\begin{eqnarray}\label{FD_06_2}
  H_{ab}^{\lbrace{2,0\rbrace}(3)} & \equiv  \frac{1}{
\sqrt{-g}}\frac{\delta}{\delta g^{ab}} \int_{\cal M} d^D x\sqrt{-g}
~R_{pq} \Box R^{pq}  \nonumber \\
&   = (\Box R)_{;ab} -  \Box \Box R_{a b} +  R_{;p (a}
R^{p}_{\phantom{p}b)} -2 \, R^{pq} R_{pq;(a b)}
\nonumber \\
& \quad  +6\, R^{pq} R_{p (a ; b)q}  -2 \, (\Box R^{pq})R_{p a q b}
+
4 \, R^{pq;r}_{\phantom{pq;r} (a} R_{|rqp| b)} \nonumber \\
&  \quad +3\, R_{;p } R^p_{\phantom{p} (a;b)} -
R^{pq}_{\phantom{pq};a}R_{pq;b}
   + 2 \, R^{pq}_{\phantom{pq} ; (a} R_{b) p;q} \nonumber \\
&  \quad
    + 4 \, R^{pq;r} R_{rqp (a;b) }  -2 \, R^{pq}R_{p a}R_{q b} +4\, R^{pr}R^q_{\phantom{q}
r}R_{p a q b} \nonumber \\
&  \quad + 2\, R^{pq}R^r_{\phantom{r} (a}R_{ |rqp|  b) }
  \nonumber \\
&  \qquad \quad + g_{ab} [ -(1/2) \, \Box \Box R     -2\,  R_{;p q}
R^{pq} +  R_{pq} \Box R^{pq} \nonumber \\
&  \quad + 2\, R_{pq ; rs}R^{prqs} -(1/2)\, R_{;p}R^{;p} + (5/2) \,
R_{pq;r} R^{pq;r} \nonumber \\
&  \quad -4 \, R_{pq;r} R^{pr;q}
      -2\, R_{pq} R^{p}_{\phantom{p} r}R^{qr}  + 2\,
R_{pq}R_{rs}R^{prqs} ] ,
\end{eqnarray}

\begin{eqnarray}\label{FD_06_3}
  H_{ab}^{(6,3)(1)} &\equiv \frac{1}{
\sqrt{-g}}\frac{\delta}{\delta g^{ab}} \int_{\cal M} d^D x\sqrt{-g}
~R^3  \nonumber \\
&   = 6\, R R_{;a b}  + 6 \, R_{;a} R_{;b}  -3\, R^2 R_{ab}  \nonumber \\
&  \qquad \quad + g_{ab} [ -6\, R\Box R -6\,  R_{;p}R^{;p} + (1/2)
\, R^3   ],
\end{eqnarray}

\begin{eqnarray}\label{FD_06_4}
H_{ab}^{(6,3)(2)} &\equiv \frac{1}{ \sqrt{-g}}\frac{\delta}{\delta
g^{ab}} \int_{\cal M} d^D x\sqrt{-g}
~RR_{pq} R^{pq}  \nonumber \\
&    = R R_{;a b} -  (\Box R) R_{a b} + 2\,  R_{;p (a}
R^{p}_{\phantom{p}b)} - R \Box R_{a b} \nonumber \\
& \quad + 2\, R^{pq} R_{pq;(a b)}
  + R_{;a} R_{;b} + 2\, R_{;p } R^p_{\phantom{p} (a;b)} -2\,
R_{;p } R_{ab}^{\phantom{ab}; p} \nonumber \\
& \quad+ 2\,
R^{pq}_{\phantom{pq};a}R_{pq;b}  - R^{pq}R_{pq}R_{ab} -2\, R R^{pq}R_{p a q b} \nonumber \\
&  \qquad \quad + g_{ab} [ -(1/2) \, R\Box R -  R_{;p q}
R^{pq} -2\,  R_{pq} \Box R^{pq}  \nonumber \\
&  \quad - R_{;p}R^{;p} -2\,  R_{pq;r} R^{pq;r} +(1/2) \,   RR_{pq}
R^{pq} ],
\end{eqnarray}

\begin{eqnarray}\label{FD_06_5}
  H_{ab}^{(6,3)(3)} & \equiv \frac{1}{
\sqrt{-g}}\frac{\delta}{\delta g^{ab}} \int_{\cal M} d^D x\sqrt{-g}
~R_{pq} R^{p}_{\phantom{p}
r}R^{qr}   \nonumber \\
&     = (3/2)\,  R_{;p (a} R^{p}_{\phantom{p}b)}  -3\,  R_{p (a}
\Box R^p_{\phantom{p} b)}
  + 3\, R^{pq} R_{p (a ; b)q}
   \nonumber \\
& \quad + (3/2) \, R_{;p } R^p_{\phantom{p} (a;b)} + 3 \,
R^{pq}_{\phantom{pq} ; (a} R_{b) p;q}  -3 \, R^p_{\phantom{p} a;q}
R_{p b}^{\phantom{p b};q}
\nonumber \\
&  \quad + 3\, R^{pq}R^r_{\phantom{r} (a}R_{ |rqp|  b) } \nonumber \\
& \qquad \quad + g_{ab} [ -(3/2) \,  R_{;p q} R^{pq}  -(3/8) \,
R_{;p}R^{;p} \nonumber \\
&  \quad- (3/2) \, R_{pq;r} R^{pr;q}  - R_{pq} R^{p}_{\phantom{p}
r}R^{qr}  + (3/2) \, R_{pq}R_{rs}R^{prqs} ], \nonumber \\
&
\end{eqnarray}

\begin{eqnarray}\label{FD_06_6}
  H_{ab}^{(6,3)(4)} & \equiv \frac{1}{ \sqrt{-g}}\frac{\delta}{\delta
g^{ab}} \int_{\cal M} d^D x\sqrt{-g}
~R_{pq}R_{rs}R^{prqs}  \nonumber \\
&     = -(1/2) \, (\Box R) R_{a b} +   R_{;p (a}
R^{p}_{\phantom{p}b)}  + R^{pq} R_{pq;(a b)}
  \nonumber \\
&  \quad -2\,  R^{pq} R_{ab; pq}  - (\Box R^{pq})R_{p a q b} -2\,
R^{pq;r}_{\phantom{pq;r} (a} R_{|rqp| b)}
 \nonumber \\
& \quad + (1/4) \, R_{;a} R_{;b} - R_{;p } R_{ab}^{\phantom{ab}; p}
 + 2\, R^{pq}_{\phantom{pq};a}R_{pq;b}
   \nonumber \\
&  \quad -2\,  R^{pq}_{\phantom{pq} ; (a} R_{b) p;q}
   + R^p_{\phantom{p} a;q} R^q_{\phantom{q} b;p}  -2\, R^{pq;r}
R_{rqp (a;b) }  \nonumber \\
& \quad -2 \, R^{pq;r} R_{p a q b;r}  + R^{pq}R_{p a}R_{q b}    -2\,
R^{pr}R^q_{\phantom{q}
r}R_{p a q b} \nonumber \\
& \quad + 2 \, R^{pq}R^r_{\phantom{r} (a}R_{ |rqp|  b) }
 + 2 \, R^{pq}R^{rs}_{\phantom{rs} pa}R_{rs q b} -2\,
  R_{pq}R^{p rs}_{\phantom{p rs}a}R^q_{\phantom{q} rs b} \nonumber \\
&\qquad \quad + g_{ab} [ (1/2) \, R_{;p q}
R^{pq} -  R_{pq} \Box R^{pq} - R_{pq ; rs}R^{prqs} \nonumber \\
&  \quad-2 \, R_{pq;r} R^{pq;r} + 2 \, R_{pq;r} R^{pr;q}
      + R_{pq} R^{p}_{\phantom{p} r}R^{qr} \nonumber \\
&  \quad -(1/2) \, R_{pq}R_{rs}R^{prqs}  ],
\end{eqnarray}

\begin{eqnarray}\label{FD_06_7}
 H_{ab}^{(6,3)(5)} &\equiv \frac{1}{
\sqrt{-g}}\frac{\delta}{\delta g^{ab}} \int_{\cal M} d^D x\sqrt{-g}
~RR_{pqrs} R^{pqrs}  \nonumber \\
&     = 2\,  R R_{;a b} -4 \, R \Box R_{a b} -4 \, R^{;pq}R_{p a q
b}    +2\, R^{pqrs} R_{pqrs ; (a b) } \nonumber \\
& \quad + 8\, R_{;p } R^p_{\phantom{p} (a;b)}  -8\, R_{;p }
R_{ab}^{\phantom{ab}; p}  + 2\, R^{pqrs}_{\phantom{pqrs};a} R_{pqrs
; b } \nonumber \\
&  \quad+ 4\, R R_{p a} R^p_{\phantom{p}b}
 -4\, R R^{pq}R_{p a q b}   -2\,  R R^{pqr}_{\phantom{pqr}a
}R_{pqr b} \nonumber \\
&  \quad -R_{ab}R^{pqrs} R_{pqrs }
\nonumber \\
& \qquad \quad + g_{ab} [ -8\, R_{pq ; rs}R^{prqs} -2\, R_{pqrs;t}
R^{pqrs;t}
    \nonumber \\
&  \quad + (1/2) \,  RR_{pqrs} R^{pqrs}
  -4\, R_{pq}R^p_{\phantom{p} rst} R^{qrst } \nonumber \\
&   \quad + 2\, R_{pqrs}R^{pquv} R^{rs}_{\phantom{rs} uv} + 8\,
R_{prqs} R^{p \phantom{u} q}_{\phantom{p} u \phantom{q} v} R^{r u s
v} ],
\nonumber \\
&
\end{eqnarray}

\begin{eqnarray}\label{FD_06_8}
  H_{ab}^{(6,3)(6)} &\equiv \frac{1}{
\sqrt{-g}}\frac{\delta}{\delta g^{ab}} \int_{\cal M} d^D x\sqrt{-g}
~R_{pq}R^p_{\phantom{p} rst} R^{qrst }  \nonumber \\
&    =   R_{;p (a} R^{p}_{\phantom{p}b)} -2\, R_{p (a} \Box
R^p_{\phantom{p} b)}
 + 2\, R^{pq} R_{p (a ; b)q}  -2\,  R^{pq} R_{ab; pq}\nonumber \\
& \quad  - R^{;pq}R_{p a q b}    + 4\, R^{p \phantom{(a};
qr}_{\phantom{p } (a} R_{|pqr| b)} + (1/2) \, R^{pqrs}
R_{pqrs ; (a b) }  \nonumber \\
& \quad + R_{;p } R^p_{\phantom{p} (a;b)} - R_{;p }
R_{ab}^{\phantom{ab}; p}
   + 2\, R^{pq}_{\phantom{pq} ; (a} R_{b) p;q}  -4\,
R^p_{\phantom{p} a;q} R_{p b}^{\phantom{p b};q} \nonumber \\
& \quad + 2\, R^p_{\phantom{p} a;q} R^q_{\phantom{q} b;p} -4\,
R^{pq;r} R_{rqp (a;b) } -2\,  R^{pq;r} R_{p a q b;r} \nonumber \\
& \quad + (1/2)\, R^{pqrs}_{\phantom{pqrs};a} R_{pqrs ; b } -
R^{pqr}_{\phantom{pqr}a;s}R_{pqr b}^{\phantom{pqr
b};s}   + 2\, R^{pq}R_{p a}R_{q b}  \nonumber \\
& \quad  -2\, R^{pr}R^q_{\phantom{q} r}R_{p a q b} +2\,
R^{pq}R^r_{\phantom{r} (a}R_{ |rqp|  b) }
\nonumber \\
& \quad -2\, R^p_{\phantom{p}  (a}R^{qrs}_{\phantom{qrs} |p| }R_{
|qrs| b) }  + 2\, R_{pq}R^{p r q
s}R_{r a s b} \nonumber \\
&   \quad -2\,  R_{pq}R^{p rs}_{\phantom{p rs}a}R^q_{\phantom{q} rs
b}  + R^{pq rs}R_{pq t a }R_{rs \phantom{t} b}^{\phantom{rs} t}
 \nonumber \\
& \quad + 4\, R^{p r q s}R^t_{\phantom{t} pq a}R_{t rs b}  -
R^{pqr}_{\phantom{pqr} s } R_{pqr t}R^{s
\phantom{a} t}_{\phantom{s} a \phantom{t} b} \nonumber \\
&  \qquad \quad + g_{ab} [ -2\, R_{pq ; rs}R^{prqs} - R_{pq;r}
R^{pq;r} +
R_{pq;r} R^{pr;q} \nonumber \\
&    \quad -(1/4)\, R_{pqrs;t} R^{pqrs;t}
       + (1/4) \, R_{pqrs}R^{pquv} R^{rs}_{\phantom{rs} uv} \nonumber \\
& \quad+ R_{prqs} R^{p \phantom{u} q}_{\phantom{p} u \phantom{q} v}
R^{r u s v} ],
\end{eqnarray}

\begin{eqnarray}\label{FD_06_9}
  H_{ab}^{(6,3)(7)} &\equiv  \frac{1}{
\sqrt{-g}}\frac{\delta}{\delta g^{ab}} \int_{\cal M} d^D x\sqrt{-g}
~R_{pqrs}R^{pquv} R^{rs}_{\phantom{rs} uv}  \nonumber \\
&    = 24\,  R^{p \phantom{(a}; qr}_{\phantom{p } (a} R_{|pqr|  b)}
-12\, R^p_{\phantom{p} a;q} R_{p b}^{\phantom{p b};q} + 12\,
R^p_{\phantom{p} a;q} R^q_{\phantom{q} b;p}  \nonumber \\
& \quad +3\, R^{pqrs}_{\phantom{pqrs};a} R_{pqrs ; b }  -6\,
R^{pqr}_{\phantom{pqr}a;s}R_{pqr b}^{\phantom{pqr b};s} \nonumber \\
& \quad -6\, R^{pq}R^{rs}_{\phantom{rs} pa}R_{rs q b} +12\, R^{p r q
s}R^t_{\phantom{t} pq a}R_{t rs b} \nonumber \\
&  \quad + g_{ab} [ (1/2) \, R_{pqrs}R^{pquv} R^{rs}_{\phantom{rs}
uv} ],
\end{eqnarray}

\begin{eqnarray}\label{FD_06_10}
  H_{ab}^{(6,3)(8)} &\equiv \frac{1}{
\sqrt{-g}}\frac{\delta}{\delta g^{ab}} \int_{\cal M} d^D x\sqrt{-g}
~R_{prqs} R^{p \phantom{u}
q}_{\phantom{p} u \phantom{q} v} R^{r u s v}  \nonumber \\
&     = (3/2) \, R^{;pq}R_{p a q b} -3 \, (\Box R^{pq})R_{p a q b}
-6\, R^{pq;r}_{\phantom{pq;r} (a} R_{|rqp| b)} \nonumber \\
&  \quad -(3/4)\, R^{pqrs} R_{pqrs ; (a b) }   + 3\,
R^{pq}_{\phantom{pq};a}R_{pq;b}
   -6\, R^{pq}_{\phantom{pq} ; (a} R_{b) p;q}   \nonumber \\
& \quad + 3\, R^p_{\phantom{p} a;q} R^q_{\phantom{q} b;p}  + 6\,
R^{pq;r} R_{rqp (a;b) }  + (3/4) \, R^{pqrs}_{\phantom{pqrs};a}
R_{pqrs ; b }  \nonumber \\
&  \quad +3\, R^{pr}R^q_{\phantom{q} r}R_{p a q b} +3\,
R^p_{\phantom{p}  (a}R^{qrs}_{\phantom{qrs} |p| }R_{ |qrs| b) } \nonumber \\
& \quad+ (3/2) \, R^{pq}R^{rs}_{\phantom{rs} pa}R_{rs q b}  -3\,
R_{pq}R^{p r q s}R_{r a s b} \nonumber \\
& \quad -3\, R_{pq}R^{p rs}_{\phantom{p rs}a}R^q_{\phantom{q} rs b}
-(3/2) \, R^{pq rs}R_{pq
t a }R_{rs \phantom{t} b}^{\phantom{rs} t} \nonumber \\
& \quad
 -9\, R^{p r q s}R^t_{\phantom{t}
pq a}R_{t rs b}  + (3/2) \, R^{pqr}_{\phantom{pqr} s } R_{pqr t}R^{s
\phantom{a} t}_{\phantom{s} a \phantom{t} b} \nonumber \\
&  \qquad \quad + g_{ab} [ (1/2) \,  R_{prqs} R^{p \phantom{u}
q}_{\phantom{p} u \phantom{q} v} R^{r u s v} ].
\end{eqnarray}

\noindent It should be here noted that the ten geometrical tensors
(\ref{FD_06_1})-(\ref{FD_06_10}) are automatically conserved due to
the invariance of the actions (\ref{actionO6_1})-(\ref{actionO6_10})
under spacetime diffeomorphisms.

From the relations (\ref{actionO6_11})-(\ref{actionO6_17}) and by
using (\ref{FD_06_1})-(\ref{FD_06_10}), we can now directly obtain
the functional derivatives with respect to the metric tensor of the
seven remaining action terms. We obtain seven conserved tensors
given by

\begin{eqnarray}\label{FD_06_11}
&  H_{ab}^{(6,1)(1)} \equiv \frac{1}{ \sqrt{-g}}\frac{\delta}{\delta
g^{ab}} \int_{\cal M} d^D x\sqrt{-g} ~\Box \Box R = 0,
\end{eqnarray}

\numparts
\begin{eqnarray}
 H_{ab}^{\lbrace{2,0\rbrace}(2)} &\equiv \frac{1}{
\sqrt{-g}}\frac{\delta}{\delta g^{ab}} \int_{\cal M} d^D x\sqrt{-g}
~R_{;p q}
R^{pq}  \nonumber \\
&  =\frac{1}{2} H_{ab}^{\lbrace{2,0\rbrace}(1)}  \label{FD_06_12a}\\
&    =  (\Box R)_{;ab} - \,  (\Box R) R_{a b} + (1/2) \, R_{;a}
R_{;b} \nonumber \\
& \qquad \quad + g_{ab} \left[ -\, \Box \Box R    - (1/4) \,
R_{;p}R^{;p}\right], \label{FD_06_12b}
\end{eqnarray}
\endnumparts

\numparts
\begin{eqnarray}
 H_{ab}^{\lbrace{2,0\rbrace}(4)} &\equiv \frac{1}{
\sqrt{-g}}\frac{\delta}{\delta g^{ab}} \int_{\cal M} d^D
x\sqrt{-g} ~R_{pq ; rs}R^{prqs}  \nonumber \\
&  =-\frac{1}{4} H_{ab}^{\lbrace{2,0\rbrace}(1)} +
H_{ab}^{\lbrace{2,0\rbrace}(3)}-H_{ab}^{(6,3)(3)}+H_{ab}^{(6,3)(4)}
 \label{FD_06_13a} \\
&     = (1/2) \, (\Box R)_{;ab} -  \Box \Box R_{a b}  + (1/2) \,
R_{;p (a} R^{p}_{\phantom{p}b)}  \nonumber \\
& \quad + 3\, R_{p (a} \Box R^p_{\phantom{p} b)} - R^{pq} R_{pq;(a
b)} + 3\, R^{pq} R_{p (a ; b)q} \nonumber \\
& \quad   -2\, R^{pq} R_{ab; pq} -3\, (\Box R^{pq})R_{p a q b}  +
2\, R^{pq;r}_{\phantom{pq;r} (a} R_{|rqp| b)}  \nonumber \\
& \quad+ (3/2) \, R_{;p } R^p_{\phantom{p} (a;b)} - R_{;p }
R_{ab}^{\phantom{ab}; p} + R^{pq}_{\phantom{pq};a}R_{pq;b}
   \nonumber \\
&  \quad -3\, R^{pq}_{\phantom{pq} ; (a} R_{b) p;q}  + 3\,
R^p_{\phantom{p} a;q} R_{p b}^{\phantom{p b};q} + R^p_{\phantom{p}
a;q} R^q_{\phantom{q} b;p}  \nonumber \\
& \quad + 2\, R^{pq;r} R_{rqp (a;b) }  -2\, R^{pq;r} R_{p a q b;r} -
R^{pq}R_{p a}R_{q b} \nonumber \\
&  \quad +2\, R^{pr}R^q_{\phantom{q} r}R_{p a q b}
+R^{pq}R^r_{\phantom{r} (a}R_{ |rqp|  b) }
\nonumber \\
& \quad   +2\, R^{pq}R^{rs}_{\phantom{rs} pa}R_{rs q b}
 -2\, R_{pq}R^{p rs}_{\phantom{p rs}a}R^q_{\phantom{q} rs b} \nonumber \\
&  \qquad \quad + g_{ab} [  R_{pq ; rs}R^{prqs} +(1/2) \,  R_{pq;r}
R^{pq;r} \nonumber \\
&  \quad-(1/2)\, R_{pq;r} R^{pr;q}  ], \label{FD_06_13b}
\end{eqnarray}
\endnumparts

\numparts
\begin{eqnarray}\label{FD_06_14}
 H_{ab}^{\lbrace{1,1\rbrace}(1)} &\equiv \frac{1}{ \sqrt{-g}}
\frac{\delta}{\delta g^{ab}} \int_{\cal M} d^D
x\sqrt{-g} ~R_{;p}R^{;p}  \nonumber \\
&  = - H_{ab}^{\lbrace{2,0\rbrace}(1)}  \label{FD_06_14a}\\
&   =  -2\, (\Box R)_{;ab} +2 \,  (\Box R) R_{a b} - R_{;a}
R_{;b} \nonumber \\
& \qquad \quad + g_{ab} \left[ 2\, \Box \Box R    + (1/2) \,
R_{;p}R^{;p}\right], \label{FD_06_14b}
\end{eqnarray}
\endnumparts

\numparts
\begin{eqnarray}
  H_{ab}^{\lbrace{1,1\rbrace}(2)} &\equiv \frac{1}{
\sqrt{-g}}\frac{\delta}{\delta g^{ab}} \int_{\cal M} d^D x\sqrt{-g}
~R_{pq;r}
R^{pq;r}  \nonumber \\
&  = - H_{ab}^{\lbrace{2,0\rbrace}(3)}  \label{FD_06_15a}\\
&   = -(\Box R)_{;ab} +  \Box \Box R_{a b} -  R_{;p (a}
R^{p}_{\phantom{p}b)} +2 \, R^{pq} R_{pq;(a b)}
\nonumber \\
& \quad  -6\, R^{pq} R_{p (a ; b)q}  +2 \, (\Box R^{pq})R_{p a q b}
-
4 \, R^{pq;r}_{\phantom{pq;r} (a} R_{|rqp| b)} \nonumber \\
&  \quad -3\, R_{;p } R^p_{\phantom{p} (a;b)} +
R^{pq}_{\phantom{pq};a}R_{pq;b}
   - 2 \, R^{pq}_{\phantom{pq} ; (a} R_{b) p;q}
    \nonumber \\
&  \quad - 4 \, R^{pq;r} R_{rqp (a;b) } +2 \, R^{pq}R_{p a}R_{q b} \nonumber \\
&  \quad -4\, R^{pr}R^q_{\phantom{q} r}R_{p a q b} - 2\,
R^{pq}R^r_{\phantom{r} (a}R_{ |rqp|  b) }
  \nonumber \\
&  \qquad \quad + g_{ab} [ (1/2) \, \Box \Box R     +2\,  R_{;p q}
R^{pq} -  R_{pq} \Box R^{pq} \nonumber \\
&  \quad - 2\, R_{pq ; rs}R^{prqs}  +(1/2)\, R_{;p}R^{;p} - (5/2) \,
R_{pq;r} R^{pq;r} \nonumber \\
&  \quad +4 \, R_{pq;r} R^{pr;q}
      +2\, R_{pq} R^{p}_{\phantom{p} r}R^{qr}  - 2\,
R_{pq}R_{rs}R^{prqs} ] , \label{FD_06_15b}
\end{eqnarray}
\endnumparts

\numparts
\begin{eqnarray}
  H_{ab}^{\lbrace{1,1\rbrace}(3)} &\equiv \frac{1}{
\sqrt{-g}}\frac{\delta}{\delta g^{ab}} \int_{\cal M} d^D
x\sqrt{-g} ~R_{pq;r} R^{pr;q}  \nonumber \\
&  =-\frac{1}{4} H_{ab}^{\lbrace{2,0\rbrace}(1)}
-H_{ab}^{(6,3)(3)}+H_{ab}^{(6,3)(4)}  \label{FD_06_16a}\\
&    = -(1/2)\, (\Box R)_{;ab} -(1/2)\,  R_{;p (a}
R^{p}_{\phantom{p}b)} +3\,  R_{p (a} \Box R^p_{\phantom{p} b)}\nonumber \\
&    \quad + R^{pq} R_{pq;(a b)}
  -3\, R^{pq} R_{p (a ; b)q}  -2\, R^{pq} R_{ab; pq} \nonumber \\
& \quad- (\Box R^{pq})R_{p a q b} -2\, R^{pq;r}_{\phantom{pq;r} (a}
R_{|rqp| b)}   -(3/2)\, R_{;p } R^p_{\phantom{p} (a;b)} \nonumber \\
& \quad - R_{;p } R_{ab}^{\phantom{ab}; p} + 2\,
R^{pq}_{\phantom{pq};a}R_{pq;b}
   -5\, R^{pq}_{\phantom{pq} ; (a} R_{b) p;q} \nonumber \\
& \quad    + 3\, R^p_{\phantom{p} a;q} R_{p b}^{\phantom{p b};q} +
R^p_{\phantom{p} a;q} R^q_{\phantom{q} b;p} -2\, R^{pq;r} R_{rqp
(a;b) } \nonumber \\
& \quad -2\,  R^{pq;r} R_{p a q b;r}  + R^{pq}R_{p a}R_{q b} -2\,
R^{pr}R^q_{\phantom{q} r}R_{p a
q b} \nonumber \\
& \quad - R^{pq}R^r_{\phantom{r} (a}R_{ |rqp|  b) }  +2\,
R^{pq}R^{rs}_{\phantom{rs} pa}R_{rs q b}
-2\, R_{pq}R^{p rs}_{\phantom{p rs}a}R^q_{\phantom{q} rs b} \nonumber \\
&  \qquad \quad + g_{ab} [ (1/2)\, \Box \Box R+2\,  R_{;p q}
R^{pq} -  R_{pq} \Box R^{pq} \nonumber \\
& \quad - R_{pq ; rs}R^{prqs}  + (1/2) \, R_{;p}R^{;p} -2\, R_{pq;r}
R^{pq;r} \nonumber \\
&  \quad + (7/2)\, R_{pq;r} R^{pr;q}   +2\, R_{pq}
R^{p}_{\phantom{p} r}R^{qr}  -2\, R_{pq}R_{rs}R^{prqs}  ], \nonumber \\
& \label{FD_06_16b}
\end{eqnarray}
\endnumparts

\numparts
\begin{eqnarray}
 H_{ab}^{\lbrace{1,1\rbrace}(4)} & \equiv \frac{1}{
\sqrt{-g}}\frac{\delta}{\delta g^{ab}} \int_{\cal M} d^D
x\sqrt{-g} ~R_{pqrs;t} R^{pqrs;t}  \nonumber \\
& =H_{ab}^{\lbrace{2,0\rbrace}(1)} -4
H_{ab}^{\lbrace{2,0\rbrace}(3)}+4H_{ab}^{(6,3)(3)}
-4H_{ab}^{(6,3)(4)}  \nonumber \\
& \quad  -2H_{ab}^{(6,3)(6)}+H_{ab}^{(6,3)(7)}+4H_{ab}^{(6,3)(8)}  \label{FD_06_17a} \\
&  = -4 H_{ab}^{\lbrace{2,0\rbrace}(4)}
-2H_{ab}^{(6,3)(6)}+H_{ab}^{(6,3)(7)}+4H_{ab}^{(6,3)(8)}  \label{FD_06_17b} \\
&    = -2\, (\Box R)_{;ab} + 4\, \Box \Box R_{a b}-4\,  R_{;p (a}
R^{p}_{\phantom{p}b)}-8\, R_{p (a} \Box R^p_{\phantom{p} b)} \nonumber \\
& \quad  + 4\, R^{pq} R_{pq;(a b)} -16\, R^{pq} R_{p (a ; b)q}  + 12
\, R^{pq} R_{ab; pq} \nonumber \\
& \quad + 8\, R^{;pq}R_{p a q b}  -32 \, R^{pq;r}_{\phantom{pq;r}
(a} R_{|rqp| b)}+16\, R^{p \phantom{(a}; qr}_{\phantom{p } (a}
R_{|pqr|
b)} \nonumber \\
& \quad -4\, R^{pqrs} R_{pqrs ; (a b) } -8\,  R_{;p }
R^p_{\phantom{p} (a;b)}  + 6\, R_{;p } R_{ab}^{\phantom{ab}; p} \nonumber \\
& \quad + 8\, R^{pq}_{\phantom{pq};a}R_{pq;b}
   -16\,  R^{pq}_{\phantom{pq} ; (a} R_{b) p;q}
   -16\, R^p_{\phantom{p} a;q} R_{p b}^{\phantom{p b};q}  \nonumber \\
& \quad + 16\, R^p_{\phantom{p} a;q} R^q_{\phantom{q} b;p}  + 24\,
R^{pq;r} R_{rqp (a;b) } +12\, R^{pq;r} R_{p a q b;r} \nonumber \\
&  \quad  + 5\, R^{pqrs}_{\phantom{pqrs};a} R_{pqrs ; b } -4\,
R^{pqr}_{\phantom{pqr}a;s}R_{pqr b}^{\phantom{pqr b};s}    + 8\,
R^{pr}R^q_{\phantom{q} r}R_{p a q b} \nonumber \\
&  \quad  -8\, R^{pq}R^r_{\phantom{r} (a}R_{ |rqp|  b) }
 + 16\, R^p_{\phantom{p}  (a}R^{qrs}_{\phantom{qrs} |p| }R_{ |qrs|
b) } \nonumber \\
& \quad -8\,  R^{pq}R^{rs}_{\phantom{rs} pa}R_{rs q b} -16\,
R_{pq}R^{p r q s}R_{r a s b} \nonumber \\
&  \quad -8\, R^{pq rs}R_{pq t a }R_{rs \phantom{t} b}^{\phantom{rs}
t}
  -32\,  R^{p r q s}R^t_{\phantom{t} pq a}R_{t rs b}  \nonumber \\
& \quad +8\, R^{pqr}_{\phantom{pqr} s } R_{pqr t}R^{s
\phantom{a} t}_{\phantom{s} a \phantom{t} b} \nonumber \\
&  \qquad \quad + g_{ab} [ (1/2) \, R_{pqrs;t} R^{pqrs;t} ].
\label{FD_06_17c}
\end{eqnarray}
\endnumparts

\noindent It should be noted that the last six expressions can be
also obtained
 by considering the
variation of the action terms on the left hand side of
(\ref{actionO6_12})-(\ref{actionO6_17}) from
(\ref{varTensMet6a})-(\ref{varTensMet6e}). Of course, the
corresponding redundant calculations are tedious ones but we have
also achieved them in order to check the validity and the internal
coherence of our results.

Finally, it seems to us interesting to provide also the functional
derivative of the alternative action term constructed from the
gravitational lagrangian $R_{pqrs} \Box R^{pqrs}$. From
(\ref{scalar_O6_comp1}) and (\ref{actionO6_13}) and by using
(\ref{FD_06_1})-(\ref{FD_06_10}) or directly by using integration by
parts, we obtain for this conserved tensor

\numparts
\begin{eqnarray}
  H_{ab}^{\lbrace{2,0\rbrace}(\mathrm{alt})} &\equiv \frac{1}{
\sqrt{-g}}\frac{\delta}{\delta
g^{ab}} \int_{\cal M} d^D x\sqrt{-g} ~R_{pqrs} \Box R^{pqrs}  \nonumber \\
&  =-H_{ab}^{\lbrace{2,0\rbrace}(1)} + 4
H_{ab}^{\lbrace{2,0\rbrace}(3)}-4H_{ab}^{(6,3)(3)}
+4H_{ab}^{(6,3)(4)}  \nonumber \\
& \quad +2H_{ab}^{(6,3)(6)}-H_{ab}^{(6,3)(7)}-4H_{ab}^{(6,3)(8)} \label{FD_06_18a} \\
&  =4H_{ab}^{\lbrace{2,0\rbrace}(4)}
+2H_{ab}^{(6,3)(6)}-H_{ab}^{(6,3)(7)}-4H_{ab}^{(6,3)(8)} \label{FD_06_18b} \\
&  =-H_{ab}^{\lbrace{1,1\rbrace}(4)}  \label{FD_06_18c} .
\end{eqnarray}
\endnumparts
It appears as the opposite of $H_{ab}^{\lbrace{1,1\rbrace}(4)}$ and
its explicit expression can be obtained directly from
(\ref{FD_06_17c}).

\subsection{Irreducible form for the cubic Lovelock tensor}

In 1971, Lovelock found the most general symmetric and conserved
tensor which is quasi-linear in the second derivatives of the metric
tensor and does not contain higher derivatives. It therefore
generalizes the Einstein tensor $R_{ab}-(1/2)Rg_{ab}$ (see
Refs.~\cite{Lovelock71,LovelockRund} for the original discussion but
also the paper by Deruelle and Madore \cite{DeruelleMadore2003} for
a historical and physical presentation of this subject). Lovelock
found moreover that this tensor can be obtained by functional
derivation with respect to the metric tensor of an action
constructed from a lagrangian which is the sum of dimensionally
extended Euler densities. The Lovelock gravitational theory is an
appealing one being free of ghosts \cite{Zwieback85,Zumino86} and is
today more particularly considered in the context of string theory
and brane cosmology.

The Lovelock lagrangian ${\cal L}_{L}$ reads
\begin{equation}\label{LovelockLagrangianTotal}
{\cal L}_{L}=\sum_{n \ge 0}^{} c_k{\cal L}_{(n)}
\end{equation}
where the $c_k$ are real arbitrary coefficients while ${\cal
L}_{(n)}$ for $n\ge 1$ is the dimensionally extended Euler density
of order $n$ in the Riemann tensor (or the
Euler-Gauss-Bonnet-Lovelock invariant of order $n$) given by
\begin{equation}\label{LovelockLagrangianOrder_n}
{\cal L}_{(n)}=\frac{1}{2^n} \delta_{r_1s_1 \dots \dots
r_ns_n}^{p_1q_1 \dots \dots p_nq_n}
R^{r_1s_1}_{\phantom{r_1s_1}p_1q_1} \dots \dots R^{r_n
s_n}_{\phantom{r_n s_n}p_n q_n}.
\end{equation}
Here $\delta_{r_1s_1 \dots \dots r_ns_n}^{p_1q_1 \dots \dots
p_nq_n}$ denotes the generalized Kronecker symbol which is totally
antisymmetric in its upper and lower indices and which can be
considered as the $n\times n$ normalized determinant
\begin{equation}\label{GeneralizedKroneckerSymb}
\delta_{r_1s_1 \dots \dots r_ns_n}^{p_1q_1 \dots \dots
p_nq_n}=\sum_{\sigma \in \Pi_{2n}} \mathrm{sign}(\sigma)
\delta^{p_1}_{\sigma(r_1)}\delta^{q_1}_{\sigma(s_1)} \dots \dots
\delta^{p_n}_{\sigma(r_n)}\delta^{q_n}_{\sigma(s_n)}.
\end{equation}
We have ${\cal L}_{(0)}=1$ by convention, ${\cal L}_{(1)}=R$ and
${\cal L}_{(2)}$ which reduces to the Gauss-Bonnet density, i.e.
\begin{equation}\label{actionLovelock_2}
{\cal L}_{(2)}=R_{pqrs}R^{pqrs}-4R_{pq}R^{pq}+R^2.
\end{equation}
The part of the Lovelock lagrangian which is cubic in the Riemann
tensor is explicitly given by
\begin{eqnarray}\label{actionLovelock_3}
& {\cal L}_{(3)}= R^3   -12 \,  RR_{pq} R^{pq} + 16 \, R_{pq}
R^{p}_{\phantom{p} r}R^{qr}  + 24 \, R_{pq}R_{rs}R^{prqs} \nonumber \\
&   \quad + 3 \,RR_{pqrs} R^{pqrs}
  -24\, R_{pq}R^p_{\phantom{p} rst} R^{qrst } + 4 \,
R_{pqrs}R^{pquv} R^{rs}_{\phantom{rs} uv} \nonumber \\
&   \quad -8\,  R_{prqs} R^{p \phantom{u} q}_{\phantom{p} u
\phantom{q} v} R^{r u s v}
\end{eqnarray}
and the formalism we have developed in the previous subsections
permits us to obtain, from the corresponding action
\begin{equation}
S_{(3)}=\int_{\cal M} d^D x\sqrt{-g(x)} ~{\cal L}_{(3)}(x),  \label{actionLovelock_1}\\
\end{equation}
the cubic part of the Lovelock tensor
\begin{equation}\label{ST_Lovelock_1} G^{(3)}_{ab
}=\frac{1}{ \sqrt{-g}} \frac{\delta S_{(3)}} {\delta g^{ab }}.
\end{equation}
With our previous notations, we have
\begin{eqnarray}\label{ST_Lovelock_2} &G^{(3)}_{ab
}=H_{ab}^{(6,3)(1)}-12 \, H_{ab}^{(6,3)(2)} +16\, H_{ab}^{(6,3)(3)}
+24\, H_{ab}^{(6,3)(4)} \nonumber\\
& \quad +3\, H_{ab}^{(6,3)(5)} -24\, H_{ab}^{(6,3)(6)}+4\,
H_{ab}^{(6,3)(7)} -8\, H_{ab}^{(6,3)(8)}
\end{eqnarray}
and more explicitly, by using (\ref{FD_06_1})-(\ref{FD_06_10}), we
obtain
\begin{eqnarray}\label{ST_Lovelock_3} &G^{(3)}_{ab}
= -3\, R^2 R_{ab} + 12\, R R_{p a} R^p_{\phantom{p}b} + 12\,
R^{pq}R_{pq}R_{ab} -24\, R^{pq}R_{p a}R_{q b} \nonumber \\
& \quad  + 12 \,R R^{pq}R_{p a q b}   -24\, R^{pr}R^q_{\phantom{q}
r}R_{p a q b} +48\, R^{pq}R^r_{\phantom{r} (a}R_{ |rqp|  b) }
\nonumber \\
& \quad  - 6\, R R^{pqr}_{\phantom{pqr}a }R_{pqr b} -3\,
R_{ab}R^{pqrs} R_{pqrs } + 24\, R^p_{\phantom{p}
(a}R^{qrs}_{\phantom{qrs} |p| }R_{|qrs| b) }\nonumber \\
& \quad  + 12\, R^{pq}R^{rs}_{\phantom{rs} pa}R_{rs q b} -24\,
R_{pq}R^{p r q
s}R_{r a s b} + 24\, R_{pq}R^{p rs}_{\phantom{p rs}a}R^q_{\phantom{q} rs b} \nonumber \\
& \quad -12\, R^{pq rs}R_{pq t a }R_{rs \phantom{t} b}^{\phantom{rs}
t}
 + 24\,  R^{p r q s}R^t_{\phantom{t}
pq a}R_{t rs b}  \nonumber \\
& \quad + 12\, R^{pqr}_{\phantom{pqr} s } R_{pqr t}R^{s \phantom{a}
t}_{\phantom{s} a \phantom{t} b} \nonumber \\
& \qquad +  g_{ab}[(1/2)\,  {\cal L}_{(3)}] .
\end{eqnarray}
This result is not a new one. In fact, it has been obtained by
M\"uller-Hoissen in Ref.~\cite{Muller-Hoissen85} much more directly.
Indeed, in order to functionally derive the Lovelock lagrangian, it
is not necessary to functionally derive independently all the
geometrical terms which compound it as we have done. Here, we
recover this result mainly in order to check our previous
calculations. It should be however noted that the comparison of
(\ref{ST_Lovelock_3}) with the result of M\"uller-Hoissen is not
immediate: Indeed, in Ref.~\cite{Muller-Hoissen85}, the FKWC-basis
is not used. But, by using (\ref{tensor_O6_comp12a}), it is easy to
put the result of M\"uller-Hoissen in our irreducible form
(\ref{ST_Lovelock_3}).

\section{Applications: Renormalization in the effective action and stress-energy tensor}

\subsection{Effective action and the Gilkey-DeWitt coefficient $a_3$}

In this section, we consider a massive scalar field $\Phi$
propagating on the $D$-dimensional curved spacetime $({\cal
M},g_{ab})$ and obeying the wave equation
\begin{equation}\label{WEQ}
\left( \Box -m^2 -\xi R \right) \Phi =0.
\end{equation}
Here $m$ is the mass of the scalar field while $\xi$ is a
dimensionless factor which accounts for the possible coupling
between this field and the gravitational background. The associated
DeWitt-Schwinger effective action $W$
\cite{DeWitt65,BirrellDavies,Avramidi_PhD,AvramidiLNP2000,DeWitt03},
which contains all the information on the ultraviolet behaviour of
the quantum theory, may be represented by the asymptotic series
\cite{DeWitt03}
\begin{eqnarray}\label{DS_EffAct}
& W=\int_{\cal M} d^D x \sqrt{-g(x)} \times \nonumber \\
& \qquad \qquad \left[\frac{1}{2 (4\pi)^{D/2}} \int_0^{+\infty}
\frac{d(is)}{(is)^{D/2+1}}e^{-im^2s} \sum_{k=0}^{+\infty} a_k(x)
(is)^k \right].
\end{eqnarray}
Here the $a_k(x)$ are the diagonal DeWitt coefficients and we have
for the four first ones \numparts
\begin{eqnarray}
a_0&=1, \label{CTSexp_a0} \\
a_1&= -(\xi -1/6) \, R, \label{CTSexp_a1} \\
a_2&= -(1/6) \, (\xi -1/5) \, \Box R + (1/2) \, (\xi -1/6)^2 \,
R^2 \nonumber \\
& \quad - (1/180) \, R_{pq}R^{pq} + (1/180) \, R_{pqrs}R^{pqrs}
\label{CTSexp_a2}
\end{eqnarray}
and
\begin{eqnarray}\label{CTSexp_a3}
a_3&= -(1/60)  \, (\xi -3/14) \, \Box \Box R     \nonumber \\
&   \quad + (1/6) \, (\xi-1/6) \,(\xi-1/5) \, R\Box R - (1/90) \,
(\xi -3/14) \,  R_{;p q}
R^{pq} \nonumber \\
&  \quad -(1/630)\,  R_{pq} \Box R^{pq} + (1/105) \, R_{pq ; rs}R^{prqs} \nonumber \\
&  \quad + (1/12) \, [\xi^2- (2/5) \, \xi +17/420] \, R_{;p}R^{;p}
-(1/2520)\, R_{pq;r} R^{pq;r} \nonumber \\
&  \quad -(1/1260)\, R_{pq;r} R^{pr;q} + (1/560) \, R_{pqrs;t}
R^{pqrs;t}
    \nonumber \\
&    \quad  - (1/6) \, (\xi-1/6)^3 \, R^3   + (1/180) \, (\xi-1/6)
\,  RR_{pq} R^{pq} \nonumber \\
&  \quad + (1/5670) \, R_{pq} R^{p}_{\phantom{p} r}R^{qr} -(1/1890)
\, R_{pq}R_{rs}R^{prqs} \nonumber \\
&  \quad -(1/180) \, (\xi-1/6) \, RR_{pqrs} R^{pqrs}
  + (1/270) \, R_{pq}R^p_{\phantom{p} rst} R^{qrst }
\nonumber \\
&   \quad  -(4/2835) \, R_{pqrs}R^{pquv} R^{rs}_{\phantom{rs} uv}
-(22/2835) \, R_{prqs} R^{p \phantom{u} q}_{\phantom{p} u
\phantom{q} v} R^{r u s v}. \nonumber \\
&
\end{eqnarray}
\endnumparts
The results we have obtained in the previous section are therefore
helpful in order to understand those of the physical aspects of the
scalar field theory which are more particularly associated with the
 coefficient $a_3$ since it is of sixth order in the derivatives of the
metric tensor. In the following subsections, we shall focus our
attention on two of them. To be more precise, it is important to
recall that the effective action (\ref{DS_EffAct}) is divergent at
the lower limit of the integral over $s$ for all the positive values
of the dimension $D$. By considering the dimensionality $D$ of
spacetime as a complex number, the effective action can be
regularized by analytic continuation and its divergent part can be
extracted coherently. In a four dimensional background, the
divergent part of the effective action is proportional to
\cite{DeWitt65,BirrellDavies,Avramidi_PhD,AvramidiLNP2000,DeWitt03}
\begin{equation}\label{divActEff_dim4}
\int_{\cal M} d^4 x \sqrt{-g(x)} \left[a_2(x)-m^2 a_1(x)+ (m^4/2)
a_0(x) \right]
\end{equation}
while its regularized part is proportional, in the large mass limit,
to \cite{Avramidi_PhD,AvramidiLNP2000}
\begin{equation}\label{ActEff_dim4}
\int_{\cal M} d^4 x \sqrt{-g(x)} a_3(x).
\end{equation}
In a six dimensional background, the divergent part of the effective
action is proportional to
\begin{eqnarray}\label{divActEff_dim6}
& \int_{\cal M} d^6 x \sqrt{-g(x)} \left[a_3(x)-m^2 a_2(x)+ (m^4/2)
a_1(x) - (m^6/6) a_0(x) \right]. \nonumber \\
&
\end{eqnarray}
In both cases, it should be noted that the global (or integrated)
Gilkey-DeWitt coefficient
\begin{equation}\label{IntDeWittCoeff_gen}
\int_{\cal M} d^D x \sqrt{-g(x)}~a_3(x)
\end{equation}
plays a central role. It is therefore necessary to have at our
disposal its explicit expression as well as the expressions of the
three first global (or integrated) DeWitt coefficients. If we now
assume that spacetime has no boundary, by using (\ref{CTSexp_a3})
and (\ref{actionO6_11})-(\ref{actionO6_17}), we can easily obtained
them. We have
\begin{equation}\label{IntDeWittCoeff_0}
\int_{\cal M} d^D x\sqrt{-g} ~a_0=\int_{\cal M} d^D x \sqrt{-g},
\end{equation}
\begin{equation}\label{IntDeWittCoeff_1}
\int_{\cal M} d^D x\sqrt{-g} ~a_1= \int_{\cal M} d^D x \sqrt{-g}
\left[ -(\xi -1/6) \, R \right]
\end{equation}
and
\begin{eqnarray}\label{IntDeWittCoeff_2}
&\int_{\cal M} d^D x\sqrt{-g} ~a_2= \int_{\cal M} d^D x \sqrt{-g}
\left[(1/2) \, (\xi -1/6)^2 \, R^2 \right. \nonumber \\
& \qquad \left. - (1/180) \, R_{pq}R^{pq} + (1/180) \,
R_{pqrs}R^{pqrs} \right]
\end{eqnarray}
and
\begin{eqnarray}\label{IntDeWittCoeff_3}
& \int_{\cal M} d^D x\sqrt{-g} ~a_3= \int_{\cal M} d^D x \sqrt{-g}
\left([(1/12)\,\xi^2- (1/30)\, \xi+1/336]
  \, R\Box R \right. \nonumber \\
&  \qquad \left. +(1/840)\,  R_{pq} \Box R^{pq} \right. \nonumber \\
&    \qquad  \left. - (1/6) \, (\xi-1/6)^3 \, R^3   + (1/180) \,
(\xi-1/6)
\,  RR_{pq} R^{pq} \right. \nonumber \\
&  \qquad \left. - (4/2835) \, R_{pq} R^{p}_{\phantom{p} r}R^{qr}
+(1/945)
\, R_{pq}R_{rs}R^{prqs} \right. \nonumber \\
&  \qquad \left. -(1/180) \, (\xi-1/6) \, RR_{pqrs} R^{pqrs}
  + (1/7560) \, R_{pq}R^p_{\phantom{p} rst} R^{qrst } \right.
\nonumber \\
&   \qquad  \left. +(17/45360) \, R_{pqrs}R^{pquv}
R^{rs}_{\phantom{rs} uv} -(1/1620) \, R_{prqs} R^{p \phantom{u}
q}_{\phantom{p} u \phantom{q} v} R^{r u s v} \right). \nonumber \\
&
\end{eqnarray}

\subsection{``Irreducible" form for
the approximated stress-energy tensor obtained from the
DeWitt-Schwinger effective action in four dimensions}

In the large mass limit of the quantized scalar field, the
renormalized effective action in four dimensions reduces to
\cite{Avramidi_PhD,AvramidiLNP2000}
\begin{eqnarray}
& W_{\mathrm{ren}}= \frac{1}{32 \pi^2 m^2}\int_{\cal M} d^4
x\sqrt{-g(x)} ~a_3(x) \label{EffAct_1}
\end{eqnarray}
and from (\ref{IntDeWittCoeff_3}) we can then write the effective
action in the form (see also
Ref.~\cite{Avramidi_PhD,AvramidiLNP2000})
\begin{eqnarray}
& W_{\mathrm{ren}}= \frac{1}{192 \pi^2 m^2}\int_{\cal M} d^4
x\sqrt{-g} ~\left([(1/2)\,\xi^2- (1/5)\, \xi+1/56]
  \, R\Box R \right. \nonumber \\
&  \qquad \left. +(1/140)\,  R_{pq} \Box R^{pq} \right. \nonumber \\
&    \qquad  \left. - (\xi-1/6)^3 \, R^3   + (1/30) \, (\xi-1/6)
\,  RR_{pq} R^{pq} \right. \nonumber \\
&  \qquad \left. - (8/945) \, R_{pq} R^{p}_{\phantom{p} r}R^{qr}
+(2/315)
\, R_{pq}R_{rs}R^{prqs} \right. \nonumber \\
&  \qquad \left. -(1/30) \, (\xi-1/6) \, RR_{pqrs} R^{pqrs}
  + (1/1260) \, R_{pq}R^p_{\phantom{p} rst} R^{qrst } \right.
\nonumber \\
&   \qquad  \left. +(17/7560) \, R_{pqrs}R^{pquv}
R^{rs}_{\phantom{rs} uv} -(1/270) \, R_{prqs} R^{p \phantom{u}
q}_{\phantom{p} u \phantom{q} v} R^{r u s v} \right). \nonumber \\
& \label{EffAct_2}
\end{eqnarray}

By functional derivation of the effective action (\ref{EffAct_2}),
we obtain an approximation for the expectation value of the
stress-energy tensor associated with the scalar field. With the
notations introduced in Section 2, we can write
\begin{eqnarray}\label{ST_EffAct 1} & \langle  ~T_{ab
} ~ \rangle_{\mathrm{ren}} =\frac{2}{ \sqrt{-g}} \frac{\delta
W_{\mathrm{ren}}} {\delta
g^{ab }} \nonumber \\
& \quad = \frac{1}{96 \pi^2 m^2}\left( [(1/2)\,\xi^2- (1/5)\,
\xi+1/56] \,
H_{ab}^{\lbrace{2,0\rbrace}(1)} \right. \nonumber\\
& \quad  \left. + (1/140) \, H_{ab}^{\lbrace{2,0\rbrace}(3)} \right. \nonumber\\
& \quad  \left. - (\xi-1/6)^3 \, H_{ab}^{(6,3)(1)} + (1/30) \,
(\xi-1/6)
\, H_{ab}^{(6,3)(2)} \right. \nonumber\\
& \quad  \left. - (8/945) \, H_{ab}^{(6,3)(3)} +(2/315) \,
H_{ab}^{(6,3)(4)} -(1/30) \, (\xi-1/6) \, H_{ab}^{(6,3)(5)} \right. \nonumber \\
& \quad  \left. + (1/1260) \, H_{ab}^{(6,3)(6)}  +(17/7560) \,
H_{ab}^{(6,3)(7)}-(1/270) \, H_{ab}^{(6,3)(8)} \right) \nonumber \\
&
\end{eqnarray}
and from the relations (\ref{FD_06_1})-(\ref{FD_06_10}), we have
then explicitly
\begin{eqnarray}\label{ST_EffAct 2}  & (96 \pi^2 m^2)   \langle  ~T_{ab
} ~  \rangle_{\mathrm{ren}} \nonumber \\
&     = [\xi^2- (2/5)\, \xi+3/70] \, (\Box R)_{;ab}
-(1/140) \,  \Box \Box R_{a b} \nonumber \\
&  \quad -6\,(\xi-1/6)[\xi^2-(1/3)\xi+1/30] R R_{;a b}
 \nonumber \\
&    \quad-(\xi-1/6)(\xi-1/5)\, (\Box R) R_{a b} + (1/15)(\xi-1/7)
\,R_{;p (a} R^{p}_{\phantom{p}b)} \nonumber \\
&  \quad + (1/10)(\xi-1/6) \, R \Box R_{a b} + (1/42) \, R_{p (a}
\Box
R^p_{\phantom{p} b)}\nonumber \\
& \quad  + (1/15)(\xi-2/7) \,R^{pq} R_{pq;(a b)}
 + (2/105) \, R^{pq} R_{p (a ; b)q} \nonumber \\
&  \quad -(1/70) \, R^{pq} R_{ab; pq}
  + (2/15)(\xi-3/14)\,R^{;pq}R_{p aq b} \nonumber \\
& \quad-(1/105)\, (\Box R^{pq})R_{p a q b}  +
(4/105)\,R^{pq;r}_{\phantom{pq;r} (a} R_{|rqp|
b)} \nonumber \\
& \quad  + (2/35)\,R^{p \phantom{(a}; qr}_{\phantom{p } (a} R_{|pqr|
b)} -(1/15)(\xi-3/14)\,R^{pqrs} R_{pqrs ; (a b)
} \nonumber \\
&  \quad -6(\xi-1/4)(\xi-1/6)^2\, R_{;a} R_{;b} -(1/5)(\xi-3/14)\,
R_{;p } R^p_{\phantom{p} (a;b)} \nonumber \\
& \quad+(1/5)(\xi-17/84)\,R_{;p }R_{ab}^{\phantom{ab}; p}
+(1/15)(\xi-1/4)\, R^{pq}_{\phantom{pq};a}R_{pq;b} \nonumber \\
& \quad
   -(1/210)\,
R^p_{\phantom{p} a;q} R_{p b}^{\phantom{p b};q} +
(1/42)\,R^p_{\phantom{p} a;q} R^q_{\phantom{q} b;p} -(1/105)\,
R^{pq;r} R_{rqp (a;b) } \nonumber \\
& \quad  -(1/70)\, R^{pq;r} R_{p a q b;r}
-(1/15)(\xi-13/56)\,R^{pqrs}_{\phantom{pqrs};a} R_{pqrs ; b }
\nonumber \\
& \quad -(1/70)\, R^{pqr}_{\phantom{pqr}a;s}R_{pqr b}^{\phantom{pqr
b};s}   + 3(\xi-1/6)^3\, R^2 R_{ab} \nonumber \\
&  \quad -(2/15)(\xi-1/6)\, R R_{p a} R^p_{\phantom{p}b}
-(1/30)(\xi-1/6)\, R^{pq}R_{pq}R_{ab} \nonumber \\
&  \quad -(2/315)\, R^{pq}R_{p a}R_{q b}
+(1/15)(\xi-1/6)\, R R^{pq}R_{p a q b} \nonumber \\
& \quad   + (1/315)\,R^{pr}R^q_{\phantom{q} r}R_{p a q b} +
(1/315)\, R^{pq}R^r_{\phantom{r} (a}R_{ |rqp|  b) }
 \nonumber \\
&  \quad + (1/15)(\xi-1/6)\, R R^{pqr}_{\phantom{pqr}a }R_{pqr b}
 + (1/30)(\xi-1/6)\,R_{ab}R^{pqrs} R_{pqrs }
\nonumber \\
& \quad -(4/315)\, R^p_{\phantom{p}  (a}R^{qrs}_{\phantom{qrs} |p|
}R_{ |qrs| b) } -(2/315)\, R^{pq}R^{rs}_{\phantom{rs} pa}R_{rs q b}
\nonumber \\
&  \quad + (4/315)\, R_{pq}R^{p r q
s}R_{r a s b} -(1/315)\, R_{pq}R^{p rs}_{\phantom{p rs}a}R^q_{\phantom{q} rs b} \nonumber \\
& \quad + (2/315)\,R^{pq rs}R_{pq t a }R_{rs \phantom{t}
b}^{\phantom{rs} t}
 +  (4/63)\, R^{p r q s}R^t_{\phantom{t}
pq a}R_{t rs b}  \nonumber \\
&  \quad -(2/315)\,  R^{pqr}_{\phantom{pqr} s } R_{pqr t}R^{s
\phantom{a} t}_{\phantom{s} a \phantom{t} b} \nonumber \\
&  \qquad
+ g_{ab} [ [-\xi^2+ (2/5)\, \xi-11/280] \, \Box \Box R     \nonumber \\
&   \quad + 6(\xi-1/6)[\xi^2- (1/3)\, \xi+1/40] \, R\Box R
-(1/30)(\xi-3/14)\, R_{;p q}
R^{pq} \nonumber \\
&  \quad -(1/15)(\xi-5/28)\,  R_{pq} \Box R^{pq} + (4/15)(\xi-1/7)\, R_{pq ; rs}R^{prqs} \nonumber \\
&  \quad + 6[\xi^3-(13/24)\xi^2+(17/180)\xi-53/10080]\,R_{;p}R^{;p}
\nonumber \\
&  \quad -(1/15)(\xi-13/56)\, R_{pq;r} R^{pq;r} -(1/420)\, R_{pq;r}
R^{pr;q} \nonumber \\
&  \quad + (1/15)(\xi-19/112)\,R_{pqrs;t} R^{pqrs;t}
     - (1/2)(\xi-1/6)^3\, R^3   \nonumber \\
&    \quad + (1/60)(\xi-1/6) \, RR_{pq} R^{pq} + (1/1890)\,R_{pq}
R^{p}_{\phantom{p} r}R^{qr}  \nonumber \\
&    \quad -(1/630)\, R_{pq}R_{rs}R^{prqs} -(1/60)(\xi-1/6) \,
RR_{pqrs} R^{pqrs}
 \nonumber \\
&   \quad  + (2/15)(\xi-1/6) \,R_{pq}R^p_{\phantom{p} rst} R^{qrst }
 \nonumber \\
&    \quad -(1/15)(\xi-47/252) \, R_{pqrs}R^{pquv}
R^{rs}_{\phantom{rs} uv}  \nonumber \\
&    \quad-(4/15)(\xi-41/252) \, R_{prqs} R^{p \phantom{u}
q}_{\phantom{p} u \phantom{q} v} R^{r u s v} ] .
\end{eqnarray}
We have therefore expressed the approximated expectation value of
the stress-energy tensor associated with the scalar field on the
FKWC-basis described in Section 2. We have a final expression which
is simplified and without any ambiguities. It should be noted that
in recent articles \cite{Matyjasek2000,Matyjasek2001,Matyjasek2006},
Matyjasek has calculated this stress-energy tensor but, being only
interested by the result in particular spacetimes, he has not
obtained a general simplified result valid in an arbitrary
background.

To conclude this subsection, we would like to emphasize some
possible other simplifications coming from ``topological" and
geometrical constraints associated with the four dimensional nature
of spacetime. In that special case, it is well-known that the Euler
number
\begin{equation} \int_{\cal M} d^4 x\sqrt{-g(x)}
~{\cal L}_{(2)}(x)
\end{equation}
where ${\cal L}_{(2)}$ is given by (\ref{actionLovelock_2}) is a
topological invariant. As a consequence, its metric variation
vanishes identically and we have
\begin{equation}\label{GB_dim4}
H_{ab}^{(4,2)(1)}-4 \, H_{ab}^{(4,2)(2)} +  H_{ab}^{(4,2)(3)}=0
\end{equation}
with
\begin{eqnarray}\label{FD_04_1}
 H_{ab}^{(4,2)(1)} &\equiv \frac{1}{ \sqrt{-g}}
\frac{\delta}{\delta g^{ab}} \int_{\cal M} d^4
x\sqrt{-g} ~R^2  \nonumber \\
&  = 2 \, R_{;ab}-2\, RR_{ab} \nonumber \\
& \qquad \quad + g_{ab}[-2 \, \Box R + (1/2) \, R^2],
\end{eqnarray}
\begin{eqnarray}\label{FD_04_2}
  H_{ab}^{(4,2)(2)} & \equiv \frac{1}{ \sqrt{-g}}
\frac{\delta}{\delta g^{ab}} \int_{\cal M} d^4
x\sqrt{-g} ~R_{pq}R^{pq}  \nonumber \\
& = R_{;ab} - \Box R_{ab} -2 \, R^{pq}R_{paqb} \nonumber \\
&  \qquad \quad + g_{ab} [ -(1/2) \, \Box R +(1/2) \, R_{pq}R^{pq}],
\end{eqnarray}
\begin{eqnarray}\label{FD_04_3}
 H_{ab}^{(4,2)(3)} &\equiv \frac{1}{ \sqrt{-g}}
\frac{\delta}{\delta g^{ab}} \int_{\cal M} d^4
x\sqrt{-g} ~R_{pqrs}R^{pqrs}  \nonumber \\
& = 2 \, R_{;ab} - 4\, \Box R_{ab} +4R^p_{\phantom{p}
 a}R_{pb}-4R^{pq}R_{paqb}-2R^{pqr}_{\phantom{pqr}a}R_{pqrb}
 \nonumber  \\
 & \qquad \quad + g_{ab} [ (1/2) \, R_{pqrs}R^{pqrs}],
\end{eqnarray}
or more explicitly,
\begin{eqnarray} \label{simplifications_4a}
& -2RR_{ab}+4R^p_{\phantom{p}
 a}R_{pb}+4R^{pq}R_{paqb}-2R^{pqr}_{\phantom{pqr}a}R_{pqrb}
 \nonumber  \\
 & \quad + \frac{1}{2} g_{ab}(R_{pqrs}R^{pqrs}-4R_{pq}R^{pq}+R^2)
 =0.
\end{eqnarray}
Moreover, in four dimensions, we have the Xu's geometrical identity
(see Ref.~\cite{XuDianyan1987} as well as the introduction of
Ref.~\cite{Harvey1995} for a simplest derivation and a clear
interpretation)
\begin{eqnarray}\label{Xu_identity1}
& R^3   -8 \,  RR_{pq} R^{pq} + 8 \, R_{pq}
R^{p}_{\phantom{p} r}R^{qr}  + 8 \, R_{pq}R_{rs}R^{prqs} \nonumber \\
&   \quad +  RR_{pqrs} R^{pqrs}
  - 4\, R_{pq}R^p_{\phantom{p} rst} R^{qrst } =0
\end{eqnarray}
which permits us to write
\begin{eqnarray}\label{simplifications_Xu} &H_{ab}^{(6,3)(1)}
-8 \, H_{ab}^{(6,3)(2)} +8\, H_{ab}^{(6,3)(3)}
+8\, H_{ab}^{(6,3)(4)} \nonumber\\
& \quad + \, H_{ab}^{(6,3)(5)} - 4\, H_{ab}^{(6,3)(6)} =0.
\end{eqnarray}
Finally, in four dimensions, the cubic Lovelock lagrangian ${\cal
L}_{(3)}$ given by (\ref{actionLovelock_3}) as well as its metric
variation, i.e. the cubic Lovelock tensor $G^{(3)}_{ab}$ given by
(\ref{ST_Lovelock_3}), vanish identically. As a consequence, we have
with (\ref{GB_dim4}) or (\ref{simplifications_4a}),
(\ref{Xu_identity1}) and (\ref{simplifications_Xu}), ${\cal
L}_{(3)}=0$ and $G^{(3)}_{ab}=0$ five new geometrical relations
which could permit us to fully simplify the expression
(\ref{ST_EffAct 2}) of the approximated stress-energy tensor. We
leave this task to the interested reader because the choice of the
(scalar and tensorial) Riemann monomials to be eliminated is a
matter of taste and depends on the problem treated as well as on the
gravitational background considered. We remark however that in the
expression (\ref{ST_EffAct 1}) of this stress-tensor, it would be
certainly interesting to drop two of the ten terms by using
(\ref{simplifications_Xu}) as well as $G^{(3)}_{ab}=0$ in the form
\begin{eqnarray}\label{simplifications_4c} &H_{ab}^{(6,3)(1)}
-12 \, H_{ab}^{(6,3)(2)} +16\, H_{ab}^{(6,3)(3)}
+24\, H_{ab}^{(6,3)(4)} \nonumber\\
& \quad +3\, H_{ab}^{(6,3)(5)} -24\, H_{ab}^{(6,3)(6)}+4\,
H_{ab}^{(6,3)(7)} -8\, H_{ab}^{(6,3)(8)} =0.
\end{eqnarray}

\subsection{Infinite counterterms appearing in the left hand side of the bare
Einstein equations in six dimensions}

As we have noted in Subsection (3.1), the divergent part of the
effective action associated with the scalar field is proportional to
(\ref{divActEff_dim6}) in a six dimensional background. It can be
removed by renormalization of the Newton's gravitational constant
and of the cosmological constant and by adding to the
Einstein-Hilbert gravitational lagrangian three counterterms of
order four ($R^2$, $R_{pq} R^{pq}$ and $R_{pqrs} R^{pqrs}$) in order
to eliminate the divergences associated with the DeWitt coefficient
$a_2$ (see (\ref{IntDeWittCoeff_2})) as well as ten counterterms of
order six ($R\Box R$, $R_{pq} \Box R^{pq}$, $R^3$, $RR_{pq} R^{pq}$,
$R_{pq} R^{p}_{\phantom{p} r}R^{qr}$, $R_{pq}R_{rs}R^{prqs}$,
$RR_{pqrs} R^{pqrs}$, $R_{pq}R^p_{\phantom{p} rst} R^{qrst }$,
$R_{pqrs}R^{pquv} R^{rs}_{\phantom{rs} uv}$, $R_{prqs} R^{p
\phantom{u} q}_{\phantom{p} u \phantom{q} v} R^{r u s v}$) in order
to eliminate the divergences associated with the Gilkey-DeWitt
coefficient $a_3$ (see (\ref{IntDeWittCoeff_3})). These last ten
counterterms induce in the bare Einstein equations a correction of
sixth order in the derivative of the metric tensor which is of the
form
\begin{eqnarray}\label{BarEE_06}
& \alpha_1
H_{ab}^{\lbrace{2,0\rbrace}(1)}  + \alpha_2 \,
H_{ab}^{\lbrace{2,0\rbrace}(3)}  + \alpha_3 \, H_{ab}^{(6,3)(1)} +
\alpha_4 \, H_{ab}^{(6,3)(2)} \nonumber\\
& + \alpha_5 \, H_{ab}^{(6,3)(3)} + \alpha_6 \, H_{ab}^{(6,3)(4)}  +
\alpha_7 \, H_{ab}^{(6,3)(5)} + \alpha_8 \, H_{ab}^{(6,3)(6)} \nonumber\\
&  + \alpha_9 \, H_{ab}^{(6,3)(7)}+ \alpha_{10} \, H_{ab}^{(6,3)(8)}
\end{eqnarray}
with the coefficients $\alpha_i$ containing terms in $1/(D-6)$ and
so diverging in the physical dimension limit. Furthermore, because
in six dimensions the Euler number
\begin{equation} \int_{\cal M} d^6 x\sqrt{-g(x)}
~{\cal L}_{(3)}(x)
\end{equation}
where ${\cal L}_{(3)}$ is given by (\ref{actionLovelock_3}) is a
topological invariant, its metric variation $G^{(3)}_{ab}$ given by
(\ref{ST_Lovelock_3}) vanishes identically. We then have here again
the constraint (\ref{simplifications_4c}) which could permit us to
eliminate one of the ten contributions of order six in the bare
Einstein equations.

\section{Concluding remarks}

The metric variations of the gravitational action terms constructed
from the curvature invariants of order six in derivatives of the
metric tensor have been already considered by numerous authors (see,
for example,
Refs.~\cite{GottloberETAL1990,LuWise1993,Matyjasek2000,Matyjasek2001,Matyjasek2006,
PiedraDeOca2007a,PiedraDeOca2007b}). However, in the works achieved
until now, all the possible simplifications due to the symmetries of
the Riemann tensor as well as to Bianchi identities have not been
systematically done. As a consequence, there did not exist until
now, in literature, explicit irreducible formulas for these
functional derivatives as it was the case for the functional
derivatives with respect to the metric tensor of the action terms
constructed from the gravitational lagrangians $\Box R$, $R^2$,
$R_{pq}R^{pq}$ and $R_{pqrs}R^{pqrs}$, i.e.  from the curvature
invariants of order four in derivatives of the metric tensor (see,
for example, Ref.~\cite{BirrellDavies}). In the present paper, by
using the results obtained by Fulling, King, Wybourne and Cummings
\cite{FKWC1992} based on group theoretical considerations, we have
solved unambiguously this problem, filling up a void in quantum
field theory in curved spacetime. We have been able to then discuss
some aspects of quantization linked to the DeWitt-Schwinger
effective action (simplified form for the renormalized effective
action in four dimensions and simplified forms for the infinite
counterterms of order six in the derivatives of the metric tensor
which must appear in the left hand side of the bare Einstein
equations in six dimensions).

We think that our results could be helpful not only in four
dimensions but also in treating some aspects of the quantum physics
of extra spatial dimensions which is currently exploding under the
impulsion of string theory. In a near future, these results could be
more particularly interesting i) in order to understand the back
reaction problem (in the large mass limit of the quantized fields)
for a wide class of metrics \cite{Matyjasek2006_private} and ii) in
order to discuss, from a general point of view, the Hadamard
renormalization of the stress-energy tensor in an arbitrary
$D$-dimensional spacetime \cite{DecaniniFolacci2005b} and to study
precisely its ambiguities.

\ack We would like to thank Jerzy Matyjasek for kind correspondence
during last summer. We are furthermore grateful to Rosalind Fiamma
for help with the English.

\section{Appendix}

\subsection{Geometrical identities between Riemann polynomials of order six and ranks zero or two}

In the present subsection of the Appendix we provide some
geometrical identities which permit us to eliminate ``alternative"
Riemann monomials of order six by expressing them in terms of
elements of the FKWC-basis. These relations are more generally
useful for calculations in two-loop quantum gravity in a four
dimensional background or for calculations in one-loop quantum
gravity in higher dimensional background. All these relations can be
derived more or less trivially from the ``symmetry" properties of
the Ricci and the Riemann tensors (pair symmetry, antisymmetry,
cyclic symmetry) \numparts
\begin{eqnarray}
& & R_{ab}=R_{ba}, \label{SymRicci} \\
& & R_{abcd}=R_{cdab}, \label{SymRiemann_1} \\
& & R_{abcd}=-R_{bacd} \quad \mathrm{and} \quad R_{abcd}=-R_{abdc},
\label{SymRiemann_2} \\
& & R_{abcd}+R_{adbc}+R_{acdb}=0, \label{SymRiemann_Cycl}
\end{eqnarray}
\endnumparts
from the Bianchi identity and its consequences obtained by
contraction of index pairs
\numparts
\begin{eqnarray}
& & R_{abcd;e}+R_{abec;d}+R_{abde;c}=0 \label{AppBianchi_1} \\
& & R_{ abcd}^{\phantom{abcd};d}= R_{ac;b}-R_{bc;a}  \label{AppBianchi_2a} \\
& &R_{ ab}^{\phantom{ab};b}= (1/2) \, R_{;a} \label{AppBianchi_2b}
\end{eqnarray}
\endnumparts
as well as from the commutation of covariant derivatives in the form
(\ref{CD_NabNabTensor}).

\bigskip

 It is of course possible to derive
numerous scalar relations between scalar Riemann monomials of order
six but, in fact, it seems to us that only five of these relations
are really important:
\begin{equation}\label{scalar_O6_comp0a}
R_{pq}R^{pr;q}_{\phantom{pr;q}r} = \frac{1}{2} \, R_{;p q}
R^{pq}+R_{pq} R^{p}_{\phantom{p} r}R^{qr}-R_{pq}R_{rs}R^{prqs}
\end{equation}
and
\begin{eqnarray}\label{scalar_O6_comp1}
&  R_{pqrs} \Box R^{pqrs} = 4 \, R_{pq ; rs}R^{prqs}
+2 \, R_{pq}R^p_{\phantom{p} rst} R^{qrst } \nonumber \\
&  \qquad - \, R_{pqrs}R^{pq}_{\phantom{pq} uv} R^{rsuv} - 4\,
R_{prqs} R^{p \phantom{u} q}_{\phantom{p} u \phantom{q} v} R^{r u s
v}
\end{eqnarray}
and
 \numparts
\begin{eqnarray}\label{scalar_O6_RCub}
& & R^{prqs}R_{pquv}R_{rs \phantom{uv}}^{\phantom{rs} uv} =
\frac{1}{2}
\, R_{pqrs}R^{pquv} R^{rs}_{\phantom{rs} uv}, \label{scalar_O6_RCub1} \\
& & R_{pq}^{\phantom{pq} rs}R^{puqv}R_{rusv} = \frac{1}{4} \,
R_{pqrs}R^{pquv} R^{rs}_{\phantom{rs} uv}, \label{scalar_O6_RCub2} \\
& &R_{prqs}R^{puqv}R_{\phantom{r} v \phantom{s} u}^{r \phantom{u} s
\phantom{v} } = -\frac{1}{4} \, R_{pqrs}R^{pquv}
R^{rs}_{\phantom{rs} uv} + R_{prqs} R^{p \phantom{u} q}_{\phantom{p}
u \phantom{q} v} R^{r u s v}. \label{scalar_O6_RCub3}
\end{eqnarray}
\endnumparts

\bigskip

Similarly, among all the tensorial relations between Riemann
monomials of order six and rank two, we have chosen to retain more
particularly the 15 following ones:
\begin{eqnarray}\label{tensor_O6_comp2}
& \Box (R_{;ab}) = (\Box R)_{;ab} + 2 \, R_{;p (a}
R^{p}_{\phantom{p}b)} \nonumber \\
& \qquad - 2\, R^{;pq}R_{p a q b} - R_{;p } R_{ab}^{\phantom{ab}; p}
+2 \, R_{;p } R^p_{\phantom{p}(a;b)}
\end{eqnarray}
and
\begin{equation}\label{tensor_O6_comp4}
 R^{pq} R_{p a ;q b}=R^{pq} R_{p a ; bq}+R^{pr}R^q_{\phantom{q} r}R_{p a q b}
+R^{pq}R^r_{\phantom{r} a}R_{rqp b}
\end{equation}
and
\begin{equation}\label{tensor_O6_comp5}
R^{p \phantom{a}; q r}_{\phantom{p } a} R_{r q p b}=\frac{1}{2} \,
R^s_{\phantom{s} a}R^{p q r}_{\phantom{p q r} s}R_{p q r b}-
\frac{1}{2} \, R^{p q}R^{rs}_{\phantom{rs} p a}R_{rs q b}
\end{equation}
and
\begin{eqnarray}\label{tensor_O6_comp14}
& & R^{pq;r}R_{raqb;p}= R^{pq;r}R_{rqpb;a} + R^{pq;r}R_{paqb;r}
\end{eqnarray}
and
\numparts
\begin{eqnarray}\label{tensor_O6_comp15}
& & R^{pqrs}_{\phantom{pqrs} ;a} R_{pqr b;s}= \frac{1}{2} \,
R^{pqrs}_{\phantom{pqrs} ;a} R_{pqrs;b}, \label{tensor_O6_comp15a}\\
& & R^{prs}_{\phantom{prs}a ;q} R^q_{\phantom{q} sr b;p}=
\frac{1}{4} \, R^{pqrs}_{\phantom{pqrs} ;a} R_{pqrs;b}
\label{tensor_O6_comp15b}
\end{eqnarray}
\endnumparts
 and
\begin{eqnarray}\label{tensor_O6_comp6}
& & R^{pqrs} R_{pqr a;b s}= \frac{1}{2} \, R^{pqrs} R_{pqrs ; a b} -
R^{pqr}_{\phantom{pqr} s } R_{pqr t}R^{s \phantom{a} t}_{\phantom{s}
a \phantom{t} b}  \nonumber \\
& & \qquad  + \frac{1}{2} \, R^{pq rs}R_{pq t a }R_{rs \phantom{t}
b}^{\phantom{rs} t}  +2 \, R^{p r q s}R^t_{\phantom{t} pq a}R_{t rs
b}
\end{eqnarray}
and
\begin{eqnarray}\label{tensor_O6_comp7}
&  R^{pqr}_{\phantom{pqr}a} \Box R_{pqr b}= -2
\,R^{pq;r}_{\phantom{pq;r} b} R_{rqp a}-2 \, R^{p \phantom{b};
qr}_{\phantom{p } b} R_{pqr a} + R^p_{\phantom{p}
b}R^{qrs}_{\phantom{qrs} p }R_{qrs a}  \nonumber \\
&  \qquad  +R^{pq}R^{rs}_{\phantom{rs} pa}R_{rs q b} - R^{pq
rs}R_{pq t a }R_{rs \phantom{t} b}^{\phantom{rs} t} - 4 \,   R^{p r
q s}R^t_{\phantom{t} pq a}R_{t rs b}
\end{eqnarray}
and
\begin{eqnarray}\label{tensor_O6_comp8}
&  R^{pq}\Box R_{p a q b}= R^{pq} R_{pq;(a b)} -2 \, R^{pq} R_{p (a
;
b)q}  + R^{pq} R_{ab; pq}  \nonumber \\
&  \qquad -2 \, R^{pq}R^r_{\phantom{r} (a}R_{|rqp| b)}  -2 \,
R^{pq}R^{rs}_{\phantom{rs} pa}R_{rs q b}  \nonumber \\
&  \qquad -2 \, R_{pq}R^{p r q s}R_{r a s b} +2 \, R_{pq}R^{p
rs}_{\phantom{p rs}a}R^q_{\phantom{q} rs b}
\end{eqnarray}
and
\begin{equation}\label{tensor_O6_comp11} R_{p q}R^{p r
s}_{\phantom{p r s}a}R^q_{\phantom{q} s r b} = R_{p q}R^{p r
s}_{\phantom{p r s}a}R^q_{\phantom{q} r s b} -\frac{1}{2} \, R^{p
q}R^{r s}_{\phantom{r s} p a}R_{r s q b}
\end{equation}
and
\numparts
\begin{eqnarray} \label{tensor2_O6_cub}
& & R^{p q r s}R^t_{\phantom{t} p q a }R_{t r s b} = \frac{1}{4} \,
R^{p q r s}R_{p q t a }R_{r s \phantom{t} b}^{\phantom{r s} t},
\label{tensor_O6_comp12b} \\
& & R^{p q r  s}R_{p q t a }R^t_{\phantom{t} r s b} = - \frac{1}{2}
\, R^{p q r s}R_{p q t a }R_{r s \phantom{t} b}^{\phantom{r s} t},
\label{tensor_O6_comp12d} \\
& & R^{p r q s}R_{p q t a }R_{r s \phantom{t} b}^{\phantom{r s} t} =
\frac{1}{2} \, R^{p q r s}R_{p q t a }R_{r s \phantom{t}
b}^{\phantom{r s} t}, \label{tensor_O6_comp12a} \\
& & R^{p r q s}R^t_{\phantom{t} p q a }R_{t s r b} = R^{p r q
s}R^t_{\phantom{t} p q a }R_{t r s b} - \frac{1}{4} \, R^{p q r
s}R_{p q t a }R_{r s \phantom{t} b}^{\phantom{r s} t},
\label{tensor_O6_comp12c} \\
& & R^{p r q  s}R_{p q t a }R^t_{\phantom{t} s r b} = \frac{1}{4} \,
R^{p q r s}R_{p q t a }R_{r s \phantom{t} b}^{\phantom{r s} t}.
\label{tensor_O6_comp12e}
\end{eqnarray}
\endnumparts
There exists in particular a lot of other relations involving terms
cubic in the Riemann tensor which are useful in calculations but
they can be obtained trivially from the five previous ones.

\subsection{Elementary variations}

In this short subsection of the Appendix, we provide a list of
relations describing the behaviour of some important geometrical
tensors in an elementary variation of the metric tensor. These
relations are useful to obtain, in Section 2, the functional
derivatives with respect to the metric tensor of the action terms
constructed from the 17 scalar Riemann monomials of order six. Apart
from two of them which we have established, these relations can be
found in Ref.~\cite{ChristensenBarth83} but we prefer to collect
them i) in order to avoid the reader having to read this reference
and ii) because in certain cases we have adopted a more practical
notation.

In the elementary variation
\begin{equation}\label{varTensMet1app}
g_{ab} \to g_{ab} + h_{ab}
\end{equation}
of the metric tensor we have
 \numparts
\begin{eqnarray}
& g^{ab} \to g^{ab} + \delta g^{ab}, \label{varTensMet3a} \\
&   \sqrt{-g} \to \sqrt{-g} + \delta (\sqrt{-g}),
\label{varTensMet3b} \\
& \Gamma_{ab}^c \to \Gamma_{ab}^c + \delta \Gamma_{ab}^c, \label{varTensMet3b_bis} \\
&  R \to
R + \delta R, \label{varTensMet3c} \\
&  \Box R \to
\Box R + \delta (\Box R), \label{varTensMet3d} \\
 &
R_{ab} \to R_{ab} + \delta R_{ab},
\label{varTensMet3e} \\
& \Box R_{ab} \to \Box R_{ab} + \delta (\Box R_{ab}),
\label{varTensMet3f} \\
 &  R_{abcd} \to R_{abcd} +
\delta R_{abcd}, \label{varTensMet3g}
\end{eqnarray}
\endnumparts
with
\numparts
\begin{eqnarray}
&  \delta g^{ab}
= -g^{pa}g^{qb}  h_{pq},   \label{varTensMet4a} \\
&   \delta(\sqrt{-g}) = \frac{1}{2} \sqrt{-g} ~h \quad \mathrm{with}
\quad h\equiv g^{pq} h_{pq}, \label{varTensMet4b} \\
&  \delta \Gamma_{ab}^c=\frac{1}{2} \left(-h_{ab}^{\phantom{ab};c}
+h^c_{\phantom{c}a;b} +h^c_{\phantom{c}b;a}\right), \label{varTensMet4b_bis} \\
& \delta R = h^{pq}_{\phantom{pq} ;pq} - \Box h -R^{pq}h_{pq}, \label{varTensMet4c} \\
&  \delta (\Box R)= \Box(h^{pq}_{\phantom{pq} ;pq})-\Box\Box h -2R^{pq;r}h_{pq;r}
- R^{pq} \Box h_{pq}- (\Box R^{pq}) h_{pq} \nonumber \\
& \quad - R^{;pq} h_{pq}-R_{;p}
h^{pq}_{\phantom{pq};q}+\frac{1}{2}R_{;p}h^{;p}
, \label{varTensMet4d} \\
 & \delta R_{ab} = \frac{1}{2} (h^p_{\phantom{p} a;bp}+ h^p_{\phantom{p} b;ap}
 -h_{;ab} -\Box h_{ab}),
\label{varTensMet4e} \\
&  \delta (\Box R_{ab})=\frac{1}{2} [\Box(h^p_{\phantom{p}
a;bp})+\Box(h^p_{\phantom{p} b;ap})-\Box (h_{;ab})-\Box \Box h_{ab}]
\nonumber \\
&\quad -R_{ab;p}h^{pq}_{\phantom{pq};q}+\frac{1}{2}R_{ab;p}h^{;p}
-R_{ab;pq}h^{pq}   \nonumber \\
& \quad - \frac{1}{2}R_{ap}(\Box h^p_{\phantom{p} b}
+h^{pq}_{\phantom{pq};bq} -h_{bq}^{\phantom{bq};pq}) -
\frac{1}{2}R_{bp}(\Box h^p_{\phantom{p} a}
+h^{pq}_{\phantom{pq};aq} -h_{aq}^{\phantom{aq};pq}) \nonumber \\
& \quad - R_{ap;q}(h_b^{\phantom{b} p;q}-h_b^{\phantom{b} q;p}+h^{
pq}_{\phantom{pq};b}) - R_{bp;q}(h_a^{\phantom{a}
p;q}-h_a^{\phantom{a} q;p}+h^{ pq}_{\phantom{pq};a}),
\label{varTensMet4f} \\
 &
\delta R_{abcd} = \frac{1}{2} (
h_{ab;dc}-h_{ab;cd}+h_{ad;bc}-h_{ac;bd}-h_{bd;ac}+h_{bc;ad} )
\nonumber \\
& \quad + R^p_{\phantom{p} bcd}h_{ap}. \label{varTensMet4g}
\end{eqnarray}
\endnumparts

In the elementary variation (\ref{varTensMet1app}) of the metric
tensor, we have moreover
\numparts
\begin{eqnarray}
& R_{;a} \to R_{;a} + \delta (R_{;a}) \label{varTensMet5a} \\
& R_{;ab} \to R_{;ab} + \delta (R_{;ab}) \label{varTensMet5b} \\
& R_{ab;c} \to R_{ab;c} + \delta (R_{ab;c}) \label{varTensMet5c} \\
& R_{ab;cd} \to R_{ab;cd} + \delta (R_{ab;cd})
\label{varTensMet5d}\\
& R_{abcd;e} \to R_{abcd;e} + \delta (R_{abcd;e})
\label{varTensMet5e}
\end{eqnarray}
\endnumparts
with \numparts
\begin{eqnarray}
& \delta (R_{;a}) = (\delta R)_{;a}\label{varTensMet6a} \\
& \delta (R_{;ab}) = (\delta R)_{;ab}-R_{;p}(\delta \Gamma_{ab}^p) \label{varTensMet6b} \\
& \delta (R_{ab;c}) = (\delta R_{ab})_{;c} - R_{pa}(\delta
\Gamma_{bc}^p) - R_{pb}(\delta
\Gamma_{ac}^p),\label{varTensMet6c} \\
& \delta (R_{ab;cd}) = (\delta R_{ab})_{;cd} - (\delta
\Gamma_{ac}^p)_{;d}R_{pb}-(\delta \Gamma_{bc}^p)_{;d}R_{pa}
-(\delta \Gamma_{cd}^p)R_{ab;p} \nonumber \\
& \quad - (\delta \Gamma_{ac}^p)R_{pb;d}-(\delta
\Gamma_{bc}^p)R_{pa;d}
- (\delta \Gamma_{ad}^p)R_{pb;c}-(\delta \Gamma_{bd}^p)R_{pa;c} \label{varTensMet6d} \\
& \delta (R_{abcd;e}) = (\delta R_{abcd})_{;e} - (\delta
\Gamma_{ae}^p)R_{pbcd} - (\delta \Gamma_{be}^p)R_{apcd} \nonumber
\\
& \quad - (\delta \Gamma_{ce}^p)R_{abpd} - (\delta
\Gamma_{de}^p)R_{abcp} . \label{varTensMet6e}
\end{eqnarray}
\endnumparts

\subsection{Standard form for the approximated stress-energy tensor associated
with massive spinor and vector fields in four dimensions}

In this last subsection of the Appendix, we briefly extend the
result obtained in Subsection (3.2) by providing, in the large mass
limit, simplifified expressions for the approximated stress-energy
tensors associated with a massive spinor field and a massive vector
field propagating in a four dimensional background. These results
have been obtained very easily and quickly from the formalism
developed in Section 2 and they explicitly prove the power of this
formalism.

In the large mass limit, the renormalized effective action of a
neutral spinor field is given by \cite{Avramidi_PhD,AvramidiLNP2000}
\begin{eqnarray}
& W^{s=1/2}_{\mathrm{ren}}= \frac{1}{192 \pi^2 m^2}\int_{\cal M} d^4
x\sqrt{-g} ~\left( -(3/280)
  \, R\Box R \right. \nonumber \\
&  \qquad \left. +(1/28)\,  R_{pq} \Box R^{pq}  +(1/864) \, R^3   - (1/180) \,  RR_{pq} R^{pq} \right. \nonumber \\
&  \qquad \left. - (25/756) \, R_{pq} R^{p}_{\phantom{p} r}R^{qr}
+(47/1260)
\, R_{pq}R_{rs}R^{prqs} \right. \nonumber \\
&  \qquad \left. -(7/1440) \,  RR_{pqrs} R^{pqrs}
  + (19/1260) \, R_{pq}R^p_{\phantom{p} rst} R^{qrst } \right.
\nonumber \\
&   \qquad  \left. +(29/7560) \, R_{pqrs}R^{pquv}
R^{rs}_{\phantom{rs} uv} -(1/108) \, R_{prqs} R^{p \phantom{u}
q}_{\phantom{p} u \phantom{q} v} R^{r u s v} \right). \nonumber \\
& \label{EffAct_2 spinor}
\end{eqnarray}
By functional derivation of this effective action, we obtain for the
associated expectation value of the stress-energy tensor the
expression (here we use the notations introduced in Section 2)
\begin{eqnarray}\label{ST_EffAct 1 spinor} & \langle  ~T^{s=1/2}_{ab
} ~ \rangle_{\mathrm{ren}} =\frac{2}{ \sqrt{-g}} \frac{\delta
W^{s=1/2}_{\mathrm{ren}}} {\delta
g^{ab }} \nonumber \\
& \quad = \frac{1}{96 \pi^2 m^2}\left( -(3/280) \,
H_{ab}^{\lbrace{2,0\rbrace}(1)} \right. \nonumber\\
& \quad  \left. + (1/28) \, H_{ab}^{\lbrace{2,0\rbrace}(3)} +(1/864)
\, H_{ab}^{(6,3)(1)} - (1/180) \,
 H_{ab}^{(6,3)(2)} \right. \nonumber\\
& \quad  \left. - (25/756) \, H_{ab}^{(6,3)(3)} + (47/1260) \,
H_{ab}^{(6,3)(4)} -(7/1440) \,  H_{ab}^{(6,3)(5)} \right. \nonumber \\
& \quad  \left. + (19/1260) \, H_{ab}^{(6,3)(6)}  +(29/7560) \,
H_{ab}^{(6,3)(7)}-(1/108) \, H_{ab}^{(6,3)(8)} \right) \nonumber \\
&
\end{eqnarray}
and from the relations (\ref{FD_06_1})-(\ref{FD_06_10}), we have
then explicitly
\begin{eqnarray}\label{ST_EffAct 2 spinor}  & (96 \pi^2 m^2)   \langle  ~T^{s=1/2}_{ab
} ~  \rangle_{\mathrm{ren}} \nonumber \\
&     =  (1/70) \, (\Box R)_{;ab}
 -(1/28) \,  \Box \Box R_{a b}  -(1/120)\, R R_{;a b}
 \nonumber \\
&    \quad +(1/120)\, (\Box R) R_{a b} +(23/840)\,R_{;p (a}
R^{p}_{\phantom{p}b)}   +(1/40) \, R \Box R_{a b}
 \nonumber \\
&  \quad +(29/420) \, R_{p (a} \Box R^p_{\phantom{p} b)}  -(19/420)
\,R^{pq} R_{pq;(a b)}
  \nonumber \\
& \quad +(61/420) \, R^{pq} R_{p (a ; b)q}  -(11/105) \, R^{pq}
R_{ab; pq}
   -(1/105)\,R^{;pq}R_{p aq b} \nonumber \\
&  \quad  -(17/210)\, (\Box R^{pq})R_{p a q b}  +(
13/105)\,R^{pq;r}_{\phantom{pq;r} (a} R_{|rqp| b)} \nonumber \\
& \quad +(16/105) \,R^{p \phantom{(a}; qr}_{\phantom{p } (a}
R_{|pqr| b)}
  +(1/210)\,R^{pqrs} R_{pqrs ; (a b) }
\nonumber \\
& \quad +(19/840) \, R_{;p } R^p_{\phantom{p} (a;b)} -(1/420)
\,R_{;p }R_{ab}^{\phantom{ab}; p}
-(1/60)\, R^{pq}_{\phantom{pq} ; (a} R_{b) p;q}\nonumber \\
& \quad
  -(1/140) \, R^p_{\phantom{p} a;q} R_{p b}^{\phantom{p b};q}
  +(3/35) \,R^p_{\phantom{p} a;q} R^q_{\phantom{q} b;p}
   -(1/21)\, R^{pq;r} R_{rqp (a;b) } \nonumber \\
& \quad  -(11/105)\, R^{pq;r} R_{p a q b;r}
+(1/420)\,R^{pqrs}_{\phantom{pqrs};a} R_{pqrs ; b }
\nonumber \\
& \quad  -(4/105)\, R^{pqr}_{\phantom{pqr}a;s}R_{pqr
b}^{\phantom{pqr b};s}
 -(1/288)\, R^2 R_{ab} \nonumber \\
&  \quad -(7/360)\, R R_{p a} R^p_{\phantom{p}b}
+(1/180)\, R^{pq}R_{pq}R_{ab} \nonumber \\
&  \quad -(1/252)\, R^{pq}R_{p a}R_{q b}
+(11/360)\, R R^{pq}R_{p a q b} \nonumber \\
& \quad  +(13/1260) \,R^{pr}R^q_{\phantom{q} r}R_{p a q b}
+(97/1260)\, R^{pq}R^r_{\phantom{r} (a}R_{ |rqp|  b) }
 \nonumber \\
&  \quad +(7/720)\, R R^{pqr}_{\phantom{pqr}a }R_{pqr b} +(7/1440)
\,R_{ab}R^{pqrs} R_{pqrs }
\nonumber \\
& \quad -(73/1260)\, R^p_{\phantom{p}  (a}R^{qrs}_{\phantom{qrs} |p|
}R_{ |qrs| b) } +(19/504)\, R^{pq}R^{rs}_{\phantom{rs} pa}R_{rs q b}
\nonumber \\
&  \quad +(73/1260) \, R_{pq}R^{p r q s}R_{r a s b}
-(97/1260)\, R_{pq}R^{p rs}_{\phantom{p rs}a}R^q_{\phantom{q} rs b} \nonumber \\
& \quad +(73/2520)\,R^{pq rs}R_{pq t a }R_{rs \phantom{t}
b}^{\phantom{rs} t}
 +(239/1260)\, R^{p r q s}R^t_{\phantom{t}pq a}R_{t rs b}  \nonumber \\
&  \quad -(73/2520)\,  R^{pqr}_{\phantom{pqr} s } R_{pqr t}R^{s
\phantom{a} t}_{\phantom{s} a \phantom{t} b} \nonumber \\
&  \qquad + g_{ab} [ (1/280) \, \Box \Box R
  -(1/240)\, R\Box R
+(1/420) \, R_{;p q}
R^{pq} \nonumber \\
&  \quad +(1/105) \,  R_{pq} \Box R^{pq} +(3/70)\, R_{pq ;
rs}R^{prqs}  -(1/672)\,R_{;p}R^{;p}
\nonumber \\
&  \quad +(3/280)\, R_{pq;r} R^{pq;r} -(1/280)\, R_{pq;r} R^{pr;q}
+(1/168)\,R_{pqrs;t} R^{pqrs;t}
\nonumber \\
&  \quad +(1/1728) \, R^3
 -(1/360)\, RR_{pq} R^{pq}
-(1/945) \,R_{pq}
R^{p}_{\phantom{p} r}R^{qr}  \nonumber \\
&    \quad +(1/315)\, R_{pq}R_{rs}R^{prqs} -(7/2880)\, RR_{pqrs}
R^{pqrs}
 \nonumber \\
&   \quad
 +(7/360)\,R_{pq}R^p_{\phantom{p} rst} R^{qrst }
 -(61/15120)\, R_{pqrs}R^{pquv}
R^{rs}_{\phantom{rs} uv}  \nonumber \\
&    \quad
 -(43/1512)\, R_{prqs} R^{p \phantom{u} q}_{\phantom{p} u
\phantom{q} v} R^{r u s v} ] .
\end{eqnarray}

In the large mass limit, the renormalized effective action of a
vector field is given by \cite{Avramidi_PhD,AvramidiLNP2000}
\begin{eqnarray}
& W^{s=1}_{\mathrm{ren}}= \frac{1}{192 \pi^2 m^2}\int_{\cal M} d^4
x\sqrt{-g} ~\left(-(27/280)
  \, R\Box R \right. \nonumber \\
&  \qquad \left. +(9/28)\,  R_{pq} \Box R^{pq}  - (5/72) \, R^3   + (31/60) \,   RR_{pq} R^{pq} \right. \nonumber \\
&  \qquad \left. - (52/63) \, R_{pq} R^{p}_{\phantom{p} r}R^{qr}
-(19/105)
\, R_{pq}R_{rs}R^{prqs} \right. \nonumber \\
&  \qquad \left. -(1/10) \,  RR_{pqrs} R^{pqrs}
  + (61/140) \, R_{pq}R^p_{\phantom{p} rst} R^{qrst } \right.
\nonumber \\
&   \qquad  \left. -(67/2520) \, R_{pqrs}R^{pquv}
R^{rs}_{\phantom{rs} uv} +(1/18) \, R_{prqs} R^{p \phantom{u}
q}_{\phantom{p} u \phantom{q} v} R^{r u s v} \right). \nonumber \\
& \label{EffAct_2 vector}
\end{eqnarray}
By functional derivation of this effective action, we obtain for the
associated expectation value of the stress-energy tensor the
expression (here we use again the notations introduced in Section 2)
\begin{eqnarray}\label{ST_EffAct 1 vector} & \langle  ~T^{s=1}_{ab
} ~ \rangle_{\mathrm{ren}} =\frac{2}{ \sqrt{-g}} \frac{\delta
W^{s=1}_{\mathrm{ren}}} {\delta
g^{ab }} \nonumber \\
& \quad = \frac{1}{96 \pi^2 m^2}\left( -(27/280) \,
H_{ab}^{\lbrace{2,0\rbrace}(1)} \right. \nonumber\\
& \quad  \left. + (9/28) \, H_{ab}^{\lbrace{2,0\rbrace}(3)}  -
(5/72) \, H_{ab}^{(6,3)(1)} + (31/60)
\, H_{ab}^{(6,3)(2)} \right. \nonumber\\
& \quad  \left. - (52/63) \, H_{ab}^{(6,3)(3)} -(19/105) \,
H_{ab}^{(6,3)(4)} -(1/10) \,  H_{ab}^{(6,3)(5)} \right. \nonumber \\
& \quad  \left. + (61/140) \, H_{ab}^{(6,3)(6)}  -(67/2520) \,
H_{ab}^{(6,3)(7)}+(1/18) \, H_{ab}^{(6,3)(8)} \right) \nonumber \\
&
\end{eqnarray}
and from the relations (\ref{FD_06_1})-(\ref{FD_06_10}), we have
then explicitly
\begin{eqnarray}\label{ST_EffAct 2 vector}  & (96 \pi^2 m^2)   \langle  ~T^{s=1}_{ab
} ~  \rangle_{\mathrm{ren}} \nonumber \\
&     =  (9/70)  \, (\Box R)_{;ab}
 -(9/28) \,  \Box \Box R_{a b} -(1/10)  \, R R_{;a b} \nonumber \\
&  \quad
  -(7/30) \, (\Box R) R_{a b}
+(13/35) \,R_{;p (a} R^{p}_{\phantom{p}b)}  -(7/60)  \, R \Box R_{a
b}
 \nonumber \\
&  \quad  +(337/210) \, R_{p (a} \Box R^p_{\phantom{p} b)} +(22/105)
\,R^{pq} R_{pq;(a b)}
  \nonumber \\
&  \quad +(34/105)  \, R^{pq} R_{p (a ; b)q}  -(107/210)  \, R^{pq}
R_{ab; pq}
   +(1/21) \,R^{;pq}R_{p aq b} \nonumber \\
& \quad  -(22/35) \, (\Box R^{pq})R_{p a q b} +(46/35)
\,R^{pq;r}_{\phantom{pq;r} (a}
R_{|rqp| b)} \nonumber \\
& \quad   +(116/105) \,R^{p \phantom{(a}; qr}_{\phantom{p } (a}
R_{|pqr| b)} -(1/42)   \,R^{pqrs} R_{pqrs ; (a b)
} \nonumber \\
&  \quad   -(1/24) \, R_{;a} R_{;b} +(83/210) \, R_{;p }
R^p_{\phantom{p} (a;b)}  -(41/84) \,R_{;p }R_{ab}^{\phantom{ab}; p}
\nonumber \\
& \quad +(31/60)   \, R^{pq}_{\phantom{pq};a}R_{pq;b} -(14/15) \,
R^{pq}_{\phantom{pq} ; (a} R_{b) p;q}
   +(221/210) \, R^p_{\phantom{p} a;q} R_{p b}^{\phantom{p b};q}
  \nonumber \\
& \quad +(113/210) \,R^p_{\phantom{p} a;q} R^q_{\phantom{q} b;p}
   +(5/21) \, R^{pq;r} R_{rqp (a;b) } \nonumber \\
& \quad -(107/210) \, R^{pq;r} R_{p a q b;r} -(17/840)
\,R^{pqrs}_{\phantom{pqrs};a} R_{pqrs ; b }
\nonumber \\
& \quad -(29/105)    \, R^{pqr}_{\phantom{pqr}a;s}R_{pqr
b}^{\phantom{pqr b};s} + (5/24) \, R^2 R_{ab}  -(2/5)  \, R R_{p a}
R^p_{\phantom{p}b} \nonumber \\
&  \quad -(31/60)  \, R^{pq}R_{pq}R_{ab} +(1/21)  \, R^{pq}R_{p
a}R_{q b}
-(19/30)  \, R R^{pq}R_{p a q b} \nonumber \\
& \quad   +(33/35) \,R^{pr}R^q_{\phantom{q} r}R_{p a q b} -(139/105)
\, R^{pq}R^r_{\phantom{r} (a}R_{ |rqp|  b) }
 \nonumber \\
&  \quad +(1/5) \, R R^{pqr}_{\phantom{pqr}a }R_{pqr b}
 +(1/10) \,R_{ab}R^{pqrs} R_{pqrs }
\nonumber \\
& \quad -(74/105) \, R^p_{\phantom{p}  (a}R^{qrs}_{\phantom{qrs} |p|
}R_{ |qrs| b) } -(5/42) \, R^{pq}R^{rs}_{\phantom{rs} pa}R_{rs q b}
\nonumber \\
&  \quad +(74/105) \, R_{pq}R^{p r q s}R_{r a s b}
-(71/105) \, R_{pq}R^{p rs}_{\phantom{p rs}a}R^q_{\phantom{q} rs b} \nonumber \\
& \quad +(37/105) \,R^{pq rs}R_{pq t a }R_{rs \phantom{t}
b}^{\phantom{rs} t}
 +(97/105) \, R^{p r q s}R^t_{\phantom{t}pq a}R_{t rs b}  \nonumber \\
&  \quad -(37/105)  \,  R^{pqr}_{\phantom{pqr} s } R_{pqr t}R^{s
\phantom{a} t}_{\phantom{s} a \phantom{t} b} \nonumber \\
&  \qquad
+ g_{ab} [
  (9/280) \, \Box \Box R
 +(19/120) \, R\Box R
   -(1/84) \, R_{;p q}
R^{pq} \nonumber \\
&  \quad -(223/420) \,  R_{pq} \Box R^{pq} +(79/105)   \, R_{pq ;
rs}R^{prqs} \nonumber \\
&  \quad+(163/1680)  \,R_{;p}R^{;p}
 -(17/56) \, R_{pq;r} R^{pq;r} +(11/420) \, R_{pq;r}
R^{pr;q}  \nonumber \\
&  \quad +(51/560) \,R_{pqrs;t} R^{pqrs;t} -(5/144)  \, R^3
 +(31/120) \, RR_{pq} R^{pq}
 \nonumber \\
&    \quad +(1/630) \,R_{pq} R^{p}_{\phantom{p} r}R^{qr}   -(53/105)
\, R_{pq}R_{rs}R^{prqs} \nonumber \\
&    \quad-(1/20) \, RR_{pqrs} R^{pqrs}
 +(2/5) \,R_{pq}R^p_{\phantom{p} rst} R^{qrst }
 \nonumber \\
&    \quad
 -(263/2520) \, R_{pqrs}R^{pquv}
R^{rs}_{\phantom{rs} uv}    -(106/315)
 \, R_{prqs} R^{p \phantom{u} q}_{\phantom{p} u
\phantom{q} v} R^{r u s v} ] . \nonumber \\
&
\end{eqnarray}

It should be finally noted that it is possible to simplify
(\ref{ST_EffAct 1 spinor}) and (\ref{ST_EffAct 1 vector}) or
(\ref{ST_EffAct 2 spinor}) and (\ref{ST_EffAct 2 vector}) by using
the ``topological" and geometrical constraints described at the end
of Subsection (3.2) which are independent of the quantum field
considered.

\end{document}